\documentclass[pdflatex,sn-mathphys-num]{sn-jnl}

\usepackage{graphicx}%
\usepackage{multirow}%
\usepackage{amsmath,amssymb,amsfonts}%
\usepackage{amsthm}%
\usepackage{mathrsfs}%
\usepackage[title]{appendix}%
\usepackage{xcolor}%
\usepackage{textcomp}%
\usepackage{manyfoot}%
\usepackage{booktabs}%
\usepackage{algorithm}%
\usepackage{algorithmicx}%
\usepackage{algpseudocode}%
\usepackage{listings}%

\theoremstyle{thmstyleone}%
\newtheorem{theorem}{Theorem}
\newtheorem{proposition}[theorem]{Proposition}%

\theoremstyle{thmstyletwo}%

\theoremstyle{thmstylethree}%

\begin{document}

\title[Article Title]{The \textit{Way} from Rota to Quantum Mechanics}

\author{David Ellerman}

\affil{\orgname{Independent researcher}, \orgaddress{\street{Trg Prekomorskih Birgad 7}, \city{Ljubljana}, \postcode{1000}, \country{Slovenia}}, ORCiD: {0000-0002-5718-618X}, \email{david@ellerman.org}}


\abstract{This paper traces an intellectual journey or \textit{Way} (in the sense of a
	Tao) that starts with some unfinished work of Gian-Carlo Rota on making a
	logic of equivalence relations or partitions. Rota understood the
	category-theoretic duality between subsets and partitions which implied there
	should be a logic of partitions dual to the usual Boolean logic of subsets.
	And just as probability starts quantitatively with the size of a subset, so he
	saw that information should start with some notion of size of a partition.
	After developing the logic of partitions and its quantitative version as
	logical entropy, it became clear that there is a fundamental duality, fully
	developed only in category theory, that runs through the exact sciences.
	Classical physics lies on the subset side and quantum physics on the partition
	side of the duality. The rest of the paper develops the treatment of quantum
	mechanics seen through the lens of partitions as the logic of definiteness and
	indefiniteness. The lattices of partitions allows the treatment of quantum
	phenomena in highly simplified but essential terms. Since Feynman saw the
	``only mystery'' of quantum mechanics in the
	two-slit experiment, this new approach is developed to show how to resolve
	that mystery. Finally, quantum statistics is treated using Rota-style enumerative combinatorics.}

\keywords{Gian-Carlo Rota, Subset/partitions duality, logic of partitions, logical entropy, q-analogs, Rota combinatorial theory}



\maketitle

\section{Introduction}\label{sec1}

This paper is a result of my collaboration with Gian-Carlo Rota in the
mid-80s. Rota died suddenly at the young age of 66 in 1999. Those who had
worked with him picked up unfinished threads of his work to see if we could
developed them further. My choice, which turned out to be more fruitful than
I could imagine, was his joint paper with two young colleagues on the logic of a
certain type of equivalence relations \cite{rota:logiccers}. An
\textit{equivalence relation} on a universe set $U=\left\{  u_{1},...,u_{n}\right\}  $ is a reflexive, symmetric, and transitive binary
relation on the set. Each point in the set has an equivalence class of other
points equivalent to it and the equivalence classes form a partition on the
set, i.e., a set of non-empty subsets of $U$ that are disjoint and cover all
of $U$. Thus the notions of a partition and an equivalence relation are
essentially the same mathematical concept (Rota would say ``cryptomorphic'') viewed from different viewpoints. The
viewpoint taken here will focus on the notion of partitions. In spite of the
label ``logic'', that paper used only the
lattice operations of join and meet and didn't have an operation of
implication on partitions. The lattices of partitions are so varied that the
only formulas true in all partition lattices are general lattice identities
\cite{whitman:lat} which is why the Rota paper focused on a specific type
(``commuting'') of equivalence relation.

Hence the first step of extending Rota's work on this thread had two
challenges: to develop the operation of implication for partitions and to
develop the general logic of arbitrary partitions on a universe set. This step
lead to another and then another until the result was the ``
Rota Way'' \cite{kung;rota} to quantum mechanics that is the
topic of this paper.

\section{The logic of partitions}\label{sec2}
\subsection{The subset-partition duality}\label{subsec2}

``Logic'' is usually taken to be the Boolean
logic of subsets, usually presented only in the special case of propositional
logic, and its sublogics and extensions. Rota knew that it was important to
focus on general subsets, rather than just subsets of $1$ ($1$ and $\emptyset$
representing truth and falsehood of propositions) since subset and partitions
are dual concepts in category theory. Given a set function $f:X\rightarrow Y$,
its \textit{image} $f\left(  X\right)  $ is a subset of $Y$ and its dual
\textit{coimage} (inverse image) is a partition $f^{-1}=\left\{  f^{-1}\left(
y\right)  \right\}  _{y\in f\left(  X\right)  }$ on $X$. By paying attention
to this duality, the logic of partitions can be developed in parallel to the
usual Boolean logic of subsets. 

\subsection{The Boolean lattice and algebra of subsets}\label{subsec3}
The powerset $\wp\left(  U\right)  $ is a Boolean lattice with the join as set
union, the meet as set intersection, and the partial order as set inclusion.
The lattice has a top (maximal element) $U$ and a bottom (minimal element)
$\emptyset$. The Boolean lattice become a Boolean algebra by the addition of
the implication or conditional operation: for $S,T\in\wp\left(  U\right)  $,
$S\supset T:=S^{c}\cup T$ (where $S^{c}=U-S$ is the complement of $S$). The
key relationship of the set implication is that when it equals the top, then
the partial order of inclusion holds:

\begin{center}
	$S\supset T=U$ iff (if and only if) $S\subseteq T$.
\end{center}

\subsection{The lattice and algebra of partitions}\label{subsec4}
A \textit{partition} $\pi$ on a set $U=\left\{  u_{1},...,u_{n}\right\}  $ is
a set $\pi=\left\{  B_{1},...,B_{m}\right\}  $ of non-empty subsets of $U$
called \textit{blocks} that are disjoint and their union if all of $U$ (for
our expository purposes, we stick to finite universe sets). A \textit{distinction} or
\textit{dit} of $\pi$ is an ordered pair of elements of $U$ that are in
different blocks of $\pi$, and the set of all distinctions of $\pi$ is the
ditset $\operatorname*{dit}\left(  \pi\right)  \subseteq U\times U$. An
\textit{indistinction} or \textit{indit} is an ordered pair of elements of $U$ in the same block
of $\pi$, and the set of all indistinctions of $\pi$ is the indit set
$\operatorname*{indit}\left(  \pi\right)  \subseteq U\times U$. The ditset and
indit set are complementary subsets of $U\times U$, i.e., $\operatorname*{dit}%
\left(  \pi\right)  ^{c}=\operatorname*{indit}\left(  \pi\right)  $. The indit
set $\operatorname*{indit}\left(  \pi\right)  $ is the equivalence relation
version of the partition $\pi$. The binary relations that are the ditsets of a
partition might be called \textit{partition relations} but they are also know
as \textit{apartness relations}.

Given another partition $\sigma=\left\{  C_{1},...,C_{m^{\prime}}\right\}  $,
the partition \textit{join} $\pi\vee\sigma$ is the partition whose blocks are
all the non-empty intersections $B_{j}\cap C_{j^{\prime}}$. The ditset of the
join is just the union of the ditsets, $\operatorname*{dit}\left(  \pi
\vee\sigma\right)  =\operatorname*{dit}\left(  \pi\right)  \cup
\operatorname*{dit}\left(  \sigma\right)  $. By DeMorgan's law, the indit set
of the join is the intersection of their equivalence relations,
$\operatorname*{indit}\left(  \pi\vee\sigma\right)  =\operatorname*{indit}%
\left(  \pi\right)  \cap\operatorname*{indit}\left(  \sigma\right)  $ which
tells us that the intersection of two equivalence relations is always an
equivalence relation. Hence given any subset of $U\times U$, there is always a
smallest equivalence relation containing that subset. Taking the subset to be
$\operatorname*{indit}\left(  \pi\right)  \cup\operatorname*{indit}\left(
\sigma\right)  $, the smallest equivalence relation containing it is defined
as: $\operatorname*{indit}\left(  \pi\wedge\sigma\right)  $, the indit set of
the partition \textit{meet }$\pi\wedge\sigma$. The partial order (PO) that
turns the set of partitions $\Pi\left(  U\right)  $ on $U$ into a lattice is
\textit{refinement}: $\sigma\precsim\pi$ (read $\pi$ refines $\sigma$ or
$\sigma$ is refined by $\pi$) if for every $B_{j}\in\pi$, there is a block
$C_{j^{\prime}}$ such that $B_{j}\subseteq C_{j^{\prime}}$. Intuitively, $\pi$
refines $\sigma$ means that $\pi$ could be obtained from $\sigma$ by chopping
up some blocks of $\sigma$. The refinement PO is the parallel to the inclusion
partial order of the Boolean lattice, and, in fact, refinement is equivalent
to the inclusion of ditsets:

\begin{center}
	$\sigma\precsim\pi$ iff $\operatorname*{dit}\left(  \sigma\right)
	\subseteq\operatorname*{dit}\left(  \pi\right)  $
\end{center}

\noindent This plants the hint that distinctions of a partition are parallel
or dual to the elements of a subset.

The top of the lattice of partitions $\Pi\left(  U\right)  $ on $U$ is the
\textit{discrete partition} $\mathbf{1}_{U}=\left\{  \left\{  u_{i}\right\}
\right\}  _{u_{i}\in U}$ where all the blocks are singletons, and the bottom
is the indiscrete partition $\mathbf{0}_{U}=\left\{  U\right\}  $ whose only
block is all of $U$. For $U=\left\{  a,b,c\right\}  ,$the two dual lattices,
the Boolean lattice of subset and the lattice of partitions, are illustrated
in Figure 1.%

\begin{figure}[h]
	\centering
	\includegraphics[width=0.7\linewidth]{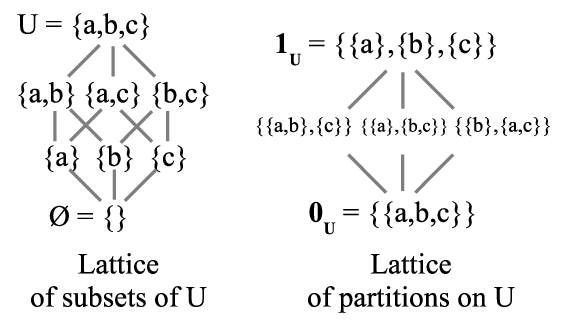}
	\caption{The two dual lattices on $U=\left\{  a,b,c\right\}  $}
	\label{fig:two-3-lattices}
\end{figure}

The lattice of partitions $\Pi\left(  U\right)  $ was known in the 19th
century (e.g., Richard Dedekind and Ernst Schr\"{o}der). But no new operations
on partitions, e.g., the implication operation, were defined in the 20th
century. In a 2001 paper commemorating Gian-Carlo Rota, the three authors
first note the fundamentality of partitions and then acknowledge the sole
operations of join and meet.

\begin{quotation}
	\noindent Equivalence relations are so ubiquitous in everyday life that we
	often forget about their proactive existence. Much is still unknown about
	equivalence relations. Were this situation remedied, the theory of equivalence
	relations could initiate a chain reaction generating new insights and
	discoveries in many fields dependent upon it.
	
	This paper springs from a simple acknowledgement: the only operations on the
	family of equivalence relations fully studied, understood and deployed are the
	binary join $\vee$ and meet $\wedge$ operations.\cite[p. 445]{bmp:eqrel}
\end{quotation}

The implication $\sigma\Rightarrow\pi$ operation on partitions can be
motivated by requiring that it satisfy the corresponding requirement as in the
Boolean algebra, i.e., the implication equals the top iff the partial order
holds, i.e., $\sigma\Rightarrow\pi=\mathbf{1}_{U}$ iff $\sigma\precsim\pi$.
The refinement $\sigma\precsim\pi$ means that for every block $B_{j}\in\pi$,
there is a block of $\sigma$ which contains it. This property of $\pi$-blocks
can be characterized by a characteristic function. That is, the implication
$\sigma\Rightarrow\pi$ can be defined simply like a characteristic or
indicator function for that inclusion-property of the $\pi$-blocks. Thus if
$B_{j}$ is contained in some $C_{j^{\prime}}\in\sigma$, then $B_{j}$ is
replaced by blocks of its discrete version, i.e., $\mathbf{1}_{B_{j}}$, the
discrete partition on $B_{j}$, and if not, then $B_{j}$ remains whole as
$B_{j}$, the block of its indiscrete version $\mathbf{0}_{B_{j}}$, the
indiscrete partition on $B_{j}$. Thus $\sigma\Rightarrow\pi$ is motivated by
seeing it as the characteristic function taking $1,0$ values according to
whether a block $B_{j}\in\pi$ is or is not contained in a block of $\sigma$.
This is how the \textit{partition implication} $\sigma\Rightarrow\pi$ is
defined and it automatically satisfies the parallel to $S\supset T=U$ iff
$S\subseteq T$:

\begin{center}
	$\sigma\Rightarrow\pi=\mathbf{1}_{U}$ iff $\sigma\precsim\pi$.
\end{center}

The addition of the partition implication operation turns the partition
lattice into the \textit{partition algebra} $\Pi\left(  U\right)  $ on $U$. In
this manner, the logic of partitions can be developed that is parallel or dual
to the Boolean logic of subsets (\cite{ell:lop}, \cite{ell:intropartitions},
\cite{ell:lop-book}).

\section{Quantitative versions of subset logic and partition logic}\label{sec3}
The next step in developing Rota's program is to define the quantitative
version of the two logics. Rota was well-aware of the category-theoretic
duality between subsets and partitions and he used it to make the key analogy:
``The lattice of partitions plays for information the role
that the Boolean algebra of subsets plays for size or
probability.'' \cite[p. 30]{kung;rota} Rota also attributed
the idea to Alfred R\'{e}nyi \cite[p. 65]{rota:fubini}. This could be
symbolized as:

\begin{center}
	$\frac{\text{Subsets}}{\text{Probability}}\approx\frac{\text{Partitions}%
	}{\text{Information}}$.
\end{center}

\noindent Probability `starts' with the probability of a subset $S$ of the set
of outcomes $U$ as the (normalized) size of the subset $\Pr\left(  S\right)
=\frac{\left\vert S\right\vert }{\left\vert U\right\vert }$. Rota reasoned
that since ``Probability is a measure on the Boolean algebra
of events'' that gives quantitatively the ``intuitive idea of the size of a set,'' we may ask by
``analogy'' for some measure
``which will capture some property that will turn out to be
for [partitions] what size is to a set.'' \cite[p.
67]{rota:fubini} Since the size of a subset is the number of elements, we
already have a hint that the dual notion to an element of a subset is a
distinction of a partition. Table 1 shows that elements and distinctions are
indeed parallel or dual notions.

\begin{center}%
	\begin{tabular}
		[c]{|l|l|}\hline
		Lattice of subsets $\wp\left(  U\right)  $ & Lattice of partitions $\Pi\left(
		U\right)  $\\\hline\hline
		Its = Elements of subsets & Dits = Distinctions of partitions\\\hline
		PO Inclusion of subsets $S\subseteq T$ & PO $\sigma\precsim\pi$ iff
		$\operatorname*{dit}\left(  \sigma\right)  \subseteq\operatorname*{dit}\left(
		\pi\right)  $\\\hline
		Join: $S\vee T=S\cup T$ & Join: $\operatorname*{dit}(\sigma\vee\pi
		)=\operatorname*{dit}\left(  \sigma\right)  \cup\operatorname*{dit}\left(
		\pi\right)  $\\\hline
		Top: $U$ all elements & Top: $\operatorname*{dit}\left(  \mathbf{1}%
		_{U}\right)  $ with all possible dits\\\hline
		Bottom: $\emptyset$ no elements & Bottom: $\operatorname*{dit}\left(
		\mathbf{0}_{U}\right)  =\emptyset$ with no dits\\\hline
	\end{tabular}

	Table 1: Duality of ``subset elements'' and
	``partition distinctions'': $\frac
	{\text{Elements}}{\text{Subsets}}\approx\frac{\text{Distinctions}%
	}{\text{Partitions}}$
\end{center}

\noindent Since no element in $U$ can be distinct from itself, i.e., cannot be
in two different blocks of a partition, the only indits of $\mathbf{1}_{U}$
are the self-pairs $\left(  u_{i},u_{i}\right)  $, i.e.,
$\operatorname*{indit}\left(  \mathbf{1}_{U}\right)  =\Delta=\left\{  \left(
u_{i},u_{i}\right)  |u_{i}\in U\right\}  $, i.e., the diagonal, so
$\operatorname*{dit}\left(  \mathbf{1}_{U}\right)  =U\times U-\Delta$ is all
possible dits. And since all the elements of $U$ are in the one block of
$\mathbf{0}_{U}$, it has no distinctions.

Rota's suggestion to measure information by the ``
size'' of a partition was correct, but he tried to shoehorn
the Shannon entropy into that role \cite[p. 68]{rota:fubini} rather than see
the duality between elements of a subset and distinctions of a partition
illustrated in Table 1. Hence Rota's suggestion can be implemented by taking
the measure of information in a partition as the (normalized) number of
distinctions in the partition so that is the beginning (equiprobable points in
$U$) definition of \textit{logical entropy}:

\begin{center}
	$h\left(  \pi\right)  =\frac{\left\vert \operatorname*{dit}\left(  \pi\right)
		\right\vert }{\left\vert U\times U\right\vert }=\frac{\left\vert U\times
		U\right\vert -\left\vert \operatorname*{indit}\left(  \pi\right)  \right\vert
	}{\left\vert U\times U\right\vert }=1-\frac{\cup_{j}\left\vert B_{j}\times
		B_{j}\right\vert }{\left\vert U\times U\right\vert }=1-\sum_{j}\left(
	\frac{\left\vert B_{j}\right\vert }{\left\vert U\right\vert }\right)  ^{2}$.
\end{center}

\noindent With the points of $U$ having probabilities $p=\left(
p_{1},...,p_{n}\right)  $, then $\Pr\left(  B_{j}\right)  =\sum_{u_{i}\in
	B_{j}}p_{i}$, and:

\begin{center}
	$h\left(  \pi\right)  =1-\sum_{j=1}^{m}\Pr\left(  B_{j}\right)  ^{2}%
	=\sum_{j\neq k}\Pr\left(  B_{j}\right)  \Pr\left(  B_{k}\right)  $
\end{center}

\noindent where the last equation holds since: $1=\left(  \sum_{j=1}^{m}%
\Pr\left(  B_{j}\right)  \right)  ^{2}$. Since the probability distribution
$p$ defines the product probability measure $p\times p$ on $U\times U$, we
also have:

\begin{center}
	$h\left(  \pi\right)  =p\times p\left(  \operatorname*{dit}\left(  \pi\right)
	\right)  $.
\end{center}

\noindent This shows the interpretation of the logical entropy $h\left(
\pi\right)  $ is the probability that in two independent draws from $U$ that
one gets a distinction of $\pi$.

Since the logical entropy is the value of a (probability) measure, the
compound notions of logical entropy are defined by the usual Venn diagram:

\begin{itemize}
	\item \textit{Joint logical entropy}: $h\left(  \pi\vee\sigma\right)  =p\times
	p\left(  \operatorname*{dit}\left(  \pi\right)  \cup\operatorname*{dit}\left(
	\sigma\right)  \right)  $,
	
	\item \textit{Difference logical entropy}: $h\left(  \pi|\sigma\right)
	=p\times p\left(  \operatorname*{dit}\left(  \pi\right)  -\operatorname*{dit}%
	\left(  \sigma\right)  \right)  $, and
	
	\item \textit{Mutual logical entropy}: $m\left(  \pi,\sigma\right)  =p\times
	p\left(  \operatorname*{dit}\left(  \pi\right)  \cap\operatorname*{dit}\left(
	\sigma\right)  \right)  $,
\end{itemize}

\noindent as illustrated in Figure 2.%

\begin{figure}[h]
	\centering
	\includegraphics[width=0.6\linewidth]{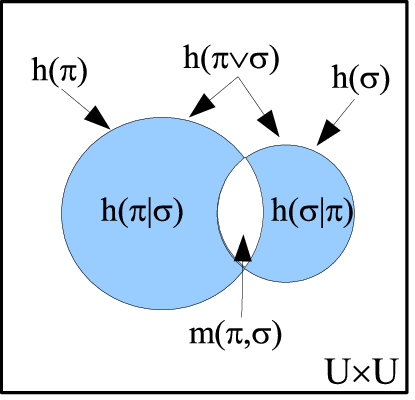}
	\caption{Compound logical entropies as values of subsets of a probability
		measure on $U\times U$}
	\label{fig:logical-venn-diagram}
\end{figure}

The logical entropy $h\left(  \mathbf{1}_{u}\right)  $ could also be taken as
just the logical entropy of the probability distribution $p$:

\begin{center}
	$h\left(  p\right)  =1-\sum_{i=1}^{n}p_{i}^{2}=\sum_{j\neq k}p_{j}p_{k}$
\end{center}

\noindent where the last equation holds since $1=\left(  \sum_{i=1}^{n}%
p_{i}\right)  ^{2}=\sum_{i=1}^{n}p_{i}^{2}+\sum_{j\neq k}p_{j}p_{k}$.

The logical entropy of a probability distribution can be illustrated in Figure
3 with a box diagram.%

\begin{figure}[h]
	\centering
	\includegraphics[width=0.7\linewidth]{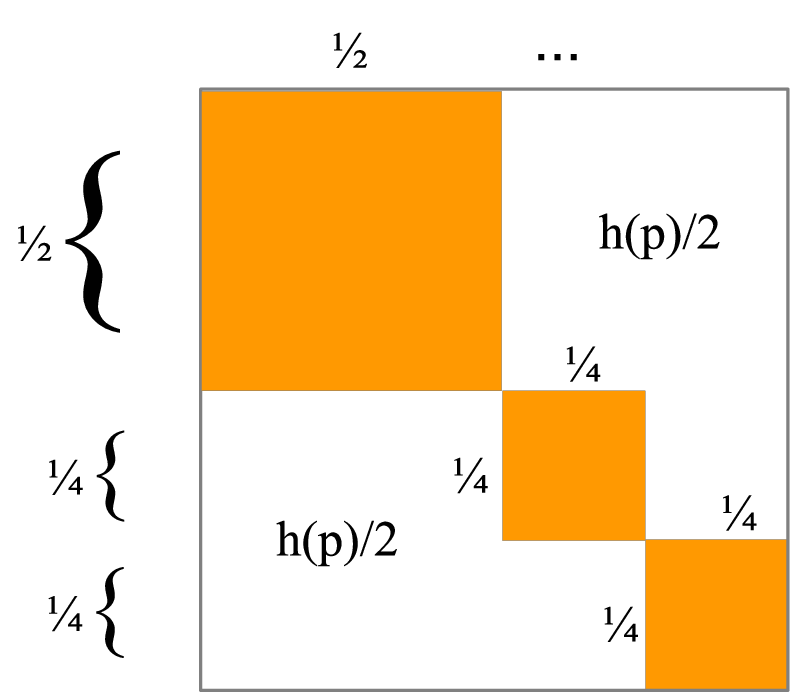}
	\caption{Logical entropy box diagram for $p=\left(  \frac{1}{2},\frac{1}%
		{4},\frac{1}{4}\right)  $}
	\label{fig:orange-log-entropy-box}
\end{figure}

\noindent For this distribution, the logical entropy is: $h\left(  p\right)
=1-\sum p_{i}^{2}=1-\left(  \frac{1}{4}+\frac{1}{16}+\frac{1}{16}\right)
=1-\frac{6}{16}=\frac{5}{8}$ with $\frac{5}{16}$ area on each side of the
diagonal squares. The derivation of the logical entropy formula as the
quantitative version of partitions is new, but the formula goes back to
Corrado Gini's index of mutability suggested in 1912 \cite{gini:vem}. Given a
probability distribution $p$ on the points of $U$ and a distance function
$d_{jk}$ between points $u_{j}$ and $u_{k}$ that is non-negative and has
$d_{ii}=0$, C. R. Rao defined the \textit{quadratic entropy} as $\sum
_{j,k}p_{j}p_{k}d_{jk}$ \cite{rao:quadentropy}. There is a \textit{logical
	distance function}, $1-\delta_{jk}$, the $1$-complement of the Kronecker delta
function, which just represent identity (distance $0$) and difference
(distance $1$). The quadratic entropy of the logical distance function is the
logical entropy: $h\left(  p\right)  =\sum_{j,k=1}^{n}p_{j}p_{k}\left(
1-\delta_{jk}\right)  $.

To summarize, the duality between subsets and partitions in their quantitative
versions gives a duality between probability theory and information theory
illustrated in Table 2.

\begin{center}%
	\begin{tabular}
		[c]{|l||l|l|}\hline
		& Logical Probability Theory & Logical Information Theory\\\hline\hline
		Outcomes & Elements of $S$ & Distinctions of $\pi$\\\hline
		Events & Subsets $S\subseteq U$ & Ditsets $\operatorname*{dit}\left(
		\pi\right)  \subseteq U\times U$\\\hline
		$p_{i}=\frac{1}{n}$ & $\Pr\left(  S\right)  =\frac{\left\vert S\right\vert
		}{\left[  U\right]  }$ & $h\left(  \pi\right)  =\frac{\left\vert
			\operatorname*{dit}\left(  \pi\right)  \right\vert }{\left\vert U\times
			U\right\vert }$\\\hline
		Probs. $p$ & $\Pr\left(  S\right)  =\sum_{u_{i}\in S}p_{i}$ & $h\left(
		\pi\right)  =\sum_{\left(  u_{i},u_{k}\right)  \in\operatorname*{dit}\left(
			\pi\right)  }p_{i}p_{k}$\\\hline
		Interpretation & $1$-draw probability of a $S$-element & $2$-draw probability
		of a $\pi$-distinction\\\hline
	\end{tabular}

	Table 2: Duality of quantitative subsets and partitions
\end{center}

\section{Introduction to the Fundamental Duality}\label{sec4}
There is a fundamental duality that starts in logic and runs throughout the
mathematical sciences \cite{ell:itsanddits}. In pure mathematics, it is the
reverse-the-arrows duality of category theory. At the logical level, it is the
category-theoretic duality of subsets and partitions. In Table 1, this duality
is extended to \textit{elements} of a subset and \textit{distinctions} of a
partition in the subset/partitions duality. Different questions arise for
single elements or pairs of elements.

\begin{itemize}
	\item For a single element, the basic questions are is it in a subset or the
	complement or does a property or its negation apply.
	
	\item For a pair of elements, the basic questions are: same or different,
	equivalent or inequivalent, indistinct or distinct, identity or difference.
\end{itemize}

Boolean subset logic applies to the single-element questions; partition logic
applies to the pair-of-elements questions. In physics, the duality separates
classical physics from quantum physics. The common-sense view of reality,
exemplified in classical physics, is that reality is definite `all the way
down.' This viewpoint is given in Leibniz's Principle of Identity of
Indistinguishables (PII) \cite[Fourth letter, p. 22]{leib-clarke:letters}.
Given two perhaps different entities, by going down deep enough, there should
be a property that applies to one and not the other; otherwise they are
identical. For Immanuel Kant, it was the Principle of Complete Determination.

\begin{quotation}
	\noindent Every thing, however, as to its possibility, further stands under
	the principle of thoroughgoing determination; according to which, among all
	possible predicates of things, insofar as they are compared with their
	opposites, one must apply to it \cite[B600]{kant:cpr}.
\end{quotation}

This characterization of the classical view of reality is quite helpful in
characterizing the quantum view of reality as \textit{not} being definite all
the way down. In quantum mechanics, there is only definiteness down to the
level given by Dirac's Complete Set of Commuting Observables (CSCO).

\begin{quotation}
	\noindent In quantum mechanics, however, identical particles are truly
	indistinguishable. This is because we cannot specify more than a complete set
	of commuting observables for each of the particles; in particular, we cannot
	label the particle by coloring it blue. \cite[p. 446]{sakurai-nap}
\end{quotation}

The definite states of a quantum particle are the eigenstates of an observable and,
beyond that, there is only the indefiniteness of particles in superposition
states. Mathematically, the linear superpositions of the definite states fill
out the vector (Hilbert) spaces that describe quantum systems. Planck's
constant $h$ is positive in the case of definiteness only down to a certain
level while we may take Planck's constant to be zero in the classical
framework. The Fundamental Duality from mathematics and logic to physics is
illustrated in Table 3.

\begin{center}%
	\begin{tabular}
		[c]{|l|c|c|}\hline
		\textbf{Fundamental Duality} & Subset side ``
		Its'' & Partition side ``
		Dits''\\\hline\hline
		{\small Math: Category Theory} &
		\multicolumn{1}{|c|}{{\small Subobjects; Limits}} & {\small Quotient objects;
			Colimits}\\\hline
		{\small Logic} & \multicolumn{1}{|c|}{{\small Logic of subsets}} &
		{\small Logic of partitions}\\\hline
		{\small Quantitative version } & \multicolumn{1}{|c|}{{\small Logical
				probability}} & {\small Logical information}\\\hline
		{\small Physics } & \multicolumn{1}{|c|}{{\small Classical physics:}} &
		{\small Quantum physics:}\\
		& \multicolumn{1}{|c|}{{\small Definite all the way down}} & {\small Definite
			to certain level; else indefinite}\\
		& \multicolumn{1}{|c|}{$h=0$} & $h>0$\\\hline
	\end{tabular}

	Table 3: Fundamental Duality from mathematics to physics
\end{center}

The logic of partitions is the logic to represent the questions applying to a
pair of elements such as definiteness versus indefiniteness. At one extreme of
the discrete partition $\mathbf{1}_{U}$, we have the full definiteness of
eigenstates so it satisfies the partition logic version of Leibniz's PII. That
is, the discrete partition makes all the distinctions that are possible so the
only indistinctions left in $\operatorname*{indit}\left(  \mathbf{1}%
_{U}\right)  $ are the self-pairs $\left(  u_{i},u_{i}\right)  $, i.e., if
$u$, $u^{\prime}\in U$ are not distinguished by the discrete partition, then
they are the same element.

\begin{center}%
	\begin{tabular}
		[c]{|c|}\hline
		If $\left(  u,u^{\prime}\right)  \in\operatorname*{indit}\left(
		\mathbf{1}_{U}\right)  $, then $u=u^{\prime}$.\\
		Partition logic PII\\\hline
	\end{tabular}

\end{center}

\noindent All the other partitions in the lattice of partitions $\Pi(U)$
contain non-singleton blocks that represent indefinite superpositions of the
definite states $u_{i}$ in the blocks.

\section{Quantum amplitudes and the Born Rule}\label{sec4}

\subsection{Redefining the sample space and events for quantum mechanics}\label{subsec5}
Rota's extensive work on probability theory included some insightful
speculations about quantum probabilities.

\begin{quotation}
	Behind the Feynman integral there lurks an even more enticing (and even less
	rigorous) concept: that of an amplitude which is meant to be the
	quantum-mechanical analog of probability (one gets probabilities by taking the
	absolute values of amplitudes and squaring them: hence the slogan
	``quantum mechanics is the imaginary square root of
	probability theory''). A concept similar to that of a sample
	space should be brought into existence for amplitudes and quantum mechanics
	should be developed starting from this concept. \cite[p. 229]{rota:fubini}
\end{quotation}

Rota's student and long-time collaborator, Henry Crapo, elaborated on this
idea as one of the `gold mines' Rota thought should be explored.

\begin{quotation}
	The second gold mine I propose from probability theory is Gian-Carlo's program
	for reshaping the fundamental concept of sample space. He approached this
	subject from several directions, recognizing in the algebra and combinatorics
	of multisets an appropriate departure from the familiar landscape of set
	theory, and recognizing in the notion of probability amplitude in quantum
	mechanics the need for a notion of sample space with more internal structure.
	It is in rewriting the foundations of quantum mechanics that new concepts of
	sample spaces would find their test of adequacy. \cite[p. 5]{crapo:goldmine}~
\end{quotation}

\noindent However, in view of the linear nature of quantum mechanics, Crapo
speculated that it ``would be natural if the new notion of
sample space pointed to by Gian-Carlo were to involve ideas from linear
algebra, \ldots\ . \cite[p. 6]{crapo:goldmine}

Indeed, it is natural that the new notion of the sample space and its events
should embody the key non-classical notion in quantum mechanics, the notion of
a superposition. The elements or outcomes in the usual sample space and its
subset events are discrete, each fully distinguished from the other outcomes.
But if we think of the appropriate corresponding quantum notion as
``the notion of `support set' of a vector relative to a
basis'' \cite[p. 6]{crapo:goldmine}, then in a superposition
of vectors in an eigenbasis, the eigenvectors are not distinguished from each
other; they are blurred or smudged together in an indefinite state. Hence we
might postulate a quantum version of the sample space and its events as having
the ``internal structure'' of the outcomes
cohering together in an indefinite state.

\subsection{Superposition events}\label{subsec6}
Let $U=\left\{  u_{1},...,u_{n}\right\}  $ be the usual sample space of
outcomes in a finite probability space with point probabilities $p$. For every
event $S\subseteq U$, we postulate a \textit{superposition event} $\Sigma S$
where the outcomes $u_{i}\in S$ are cohered together. We need to specify the
probabilities of the outcomes given a superposition event $\Sigma S$ rather
than the usual discrete event $S$. We postulate that the probabilities are the same:

\begin{center}
	$\Pr\left(  u_{i}|\Sigma S\right)  =\Pr\left(  u_{i}|S\right)  $.
\end{center}

\noindent This may seem at first a bug but it is actually a feature since the
same occurs in quantum mechanics. That is, the probabilities of a measurement
of a superposition state giving an outcome are the same as the probabilities
of a measurement of the corresponding completely decohered `classical'
statistical mixture state. The two states can only be distinguished by
measurement in another basis. Hence the probability outcomes is not how the
superposition event $\Sigma S$ differs from the classical discrete event $S$.

How then can we mathematically represent the difference between the two events?
Crapo's suggestion of linear algebra shows the way. As a vector, the classical
discrete subset-event $S$ could be represented as a $0,1$-column vector
$\left\vert S\right\rangle $ with the entries being the values of the
characteristic function $\chi_{S}:U\rightarrow\left\{  0,1\right\}  $. But
such a one-dimensional $n\times1$ vector cannot represent the cohering
together of the outcomes in the superposition event $\Sigma S$ so we need a
two-dimensional $n\times n$ matrix where the non-zero off-diagonal elements
represent the cohering together of the corresponding diagonal elements. And
the classical discrete event $S$ could be represented in the same framework by
a diagonal matrix (i.e., no non-zero off-diagonal elements) with the values
$\chi_{S}\left(  u_{i}\right)  $ down the diagonal. Ignoring probabilities for
the moment, the difference between the two events $S$ and $\Sigma S$ can be
representing using the combinatorial tool of relation matrices.

A \textit{binary relation} $R$ on $U\times U$ is a subset $R\subseteq U\times
U$. The \textit{relation matrix} (or \textit{incidence matrix})
$\operatorname*{Rel}\left(  R\right)  $ of $R$ has the entries:
$\operatorname*{Rel}\left(  R\right)  _{jk}=1$ if $\left(  u_{j},u_{k}\right)
\in R$, else $0$. Then the matrix for the classical event $S$ is the relation
matrix $\operatorname*{Rel}\left(  \Delta S\right)  $ for the diagonal $\Delta
S=\left\{  \left(  u_{i},u_{i}\right)  |u_{i}\in S\right\}  $ and the matrix
for the superposition event $\Sigma S$ is $\operatorname*{Rel}\left(  S\times
S\right)  $ where $S\times S=\left\{  \left(  u_{j},u_{k}\right)  |u_{j}%
,u_{k}\in S\right\}  $. There is the overlap in the trivial case of singletons
$S=\left\{  u_{i}\right\}  $ but, strictly speaking, a superposition
superposes two or more elements so, without loss of generality, we only
consider $\Sigma S$ for $\left\vert S\right\vert \geq2$.

The relation matrix $\operatorname*{Rel}\left(  S\times S\right)  $ associated
with a superposition event $\Sigma S$ gives the connection between partitions
and equivalence relations. Given a partition $\pi=\left\{  B_{1}%
,...,B_{m}\right\}  $, each block $B_{j}$ functions like a superposition event
and $B_{j}\times B_{j}$ is the set of indistinctions contributed to the
corresponding equivalence relation: $\operatorname*{indit}\left(  \pi\right)
=\cup_{j=1}^{m}\left(  B_{j}\times B_{j}\right)  =U\times
U-\operatorname*{dit}\left(  \pi\right)  $. Relation matrices, a notion from
combinatorial theory, may seem far removed from quantum mechanics, but the key
difference between the relation matrices $\operatorname*{Rel}\left(  \Delta
S\right)  $ and $\operatorname*{Rel}\left(  S\times S\right)  $ already
appears at this stage. The \textit{key difference} is that starting with
$\operatorname*{Rel}\left(  S\times S\right)  $ representing $\Sigma S$, it
can be obtained as an outer product $\left\vert S\right\rangle \left\vert
S\right\rangle ^{t}$ (where the superscript $t$ represents the transpose) of
the $0,1$-column vector $\left\vert S\right\rangle $ where $\left\langle
u_{i}|S\right\rangle =\chi_{S}\left(  u_{i}\right)  $ with its transpose:

\begin{center}
	$\operatorname*{Rel}\left(  S\times S\right)  =\left\vert S\right\rangle
	\left\vert S\right\rangle ^{t}$
\end{center}

\noindent whereas the relation matrix $\operatorname*{Rel}\left(  \Delta
S\right)  $ representing the classical discrete event cannot be obtained as an
outer product of a vector with its transpose (in the non-overlapping case of
$\left\vert S\right\vert \geq2$). That key difference is the `magic' that
appears already at the level of relation matrices in combinatorial theory and
eventually accounts for probability amplitudes and the Born rule.

For instance, suppose $U=\left\{  a,b,c\right\}  $ and $S=\left\{
b,c\right\}  $. Then we have:

\begin{center}
	$\left\vert S\right\rangle \left\vert S\right\rangle ^{t}=%
	\begin{bmatrix}
		0\\
		1\\
		1
	\end{bmatrix}%
	\begin{bmatrix}
		0 & 1 & 1
	\end{bmatrix}
	=%
	\begin{bmatrix}
		0 & 0 & 0\\
		0 & 1 & 1\\
		0 & 1 & 1
	\end{bmatrix}
	=\operatorname*{Rel}\left(  S\times S\right)  .$
\end{center}

Thus the superposition event $\Sigma S$, properly represented with the
non-zero off-diagonal `coherences' (i.e., indistinctions) in
$\operatorname*{Rel}\left(  S\times S\right)  $ to distinguish it from the
classical discrete event $S$ represented by $\operatorname*{Rel}\left(  \Delta
S\right)  $, brings something \textit{new}--which foreshadows probability
amplitudes. And we even have the foreshadowing of the Born Rule since the
$i^{th}$ entry in $\left\vert S\right\rangle $ times the $i^{th}$ entry in
$\left\vert S\right\rangle ^{t}$ gives the $i^{th}$ diagonal entry in
$\left\vert S\right\rangle \left\vert S\right\rangle ^{t}=\operatorname*{Rel}%
\left(  S\times S\right)  $ representing $\Sigma S$.

\subsection{From combinatorial relation matrices to quantum density matrices}\label{subsec7}

Dividing the relation matrices $\operatorname*{Rel}\left(  \Delta S\right)  $
and $\operatorname*{Rel}\left(  S\times S\right)  $ through by their trace
(sum of diagonal elements) turns them into density matrices that represent
pure quantum states:

\begin{center}
	$\rho\left(  \Sigma S\right)  =\frac{\operatorname*{Rel}(S\times
		S)}{\operatorname*{tr}\left[  \operatorname*{Rel}\left(  S\times S\right)
		\right]  }$ and $\rho\left(  S\right)  =\frac{\operatorname*{Rel}\left(
		\Delta S\right)  }{\operatorname*{tr}\left[  \operatorname*{Rel}\left(  \Delta
		S\right)  \right]  }$.
\end{center}

\noindent The trace in both cases is $\left\vert S\right\vert $ and the matrix
product $\operatorname*{Rel}\left(  S\times S\right)  \operatorname*{Rel}%
\left(  S\times S\right)  =\left\vert S\right\vert \operatorname*{Rel}\left(
S\times S\right)  $ so we have:

\begin{center}
	$\rho\left(  \Sigma S\right)  ^{2}=\frac{\operatorname*{Rel}\left(  S\times
		S\right)  }{\left\vert S\right\vert }\frac{\operatorname*{Rel}\left(  S\times
		S\right)  }{\left\vert S\right\vert }=\frac{\left\vert S\right\vert
		\operatorname*{Rel}\left(  S\times S\right)  }{\left\vert S\right\vert ^{2}%
	}=\frac{\operatorname*{Rel}\left(  S\times S\right)  }{\left\vert S\right\vert
	}=\rho\left(  \Sigma S\right)  $.
\end{center}

\noindent Hence in QM, $\rho\left(  \Sigma S\right)  $ is a pure state while
$\rho\left(  S\right)  $ ($\left\vert S\right\vert \geq2$) is a mixed state.
The eigenvalues of any density matrix are non-negative and sum to one. In the
case of a pure state density matrix, there is one eigenvalue of $1$ with the
rest being zeros. The eigenvalue of $1$ has a normalized eigenvector
$\left\vert s\right\rangle $ and since a density matrix is also Hermitian, it
has a spectral decomposition as the sum of the eigenvalues times the
projectors formed from the eigenvectors which in this case is (using the Dirac
ket-bra notation):

\begin{center}
	$\rho\left(  \Sigma S\right)  =\left\vert s\right\rangle \left\vert
	s\right\rangle ^{t}=\left\vert s\right\rangle \left\langle s\right\vert $
\end{center}

\noindent so the pure state matrix $\rho\left(  \Sigma\right)  $, like
$\operatorname*{Rel}(S\times S)$, is the outer product of a vector with its
transpose, and the mixed state $\rho\left(  S\right)  $, like
$\operatorname*{Rel}\left(  \Delta S\right)  $, is not. That vector
$\left\vert s\right\rangle $ is easily constructed (up to sign) as:
$\left\langle u_{i}|s\right\rangle =\sqrt{\frac{1}{\left\vert S\right\vert }}$
if $u_{i}\in S$, else $0$. The new vector that appears in the case of the
superposition state $\Sigma S$ (but not for the mixed classical state $S$) is
the probability amplitude. In simple terms, the probability amplitude is a
feature or consequence of superposition. And the Born rule is built into
obtaining the pure state density matrix as an outer product, i.e., $\Pr\left(
u_{i}|\Sigma S\right)  =\left\langle u_{i}|s\right\rangle ^{2}=\frac
{1}{\left\vert S\right\vert }$ if $u_{i}\in S$, else $0$.

All these relationship continue to hold as we build up to the full QM case
over the complex numbers. We might consider the next step as given by point
probabilities $p=\left(  p_{1},...,p_{n}\right)  $ on $U$. Then $\rho\left(
\Sigma S\right)  $ is the real density matrix where $\rho\left(  \Sigma
S\right)  _{jk}=\frac{\sqrt{p_{j}p_{k}}}{\Pr\left(  \Sigma S\right)  }$ if
$u_{j},u_{k}\in S$, else $0$, where $\Pr\left(  \Sigma S\right)  =\Pr\left(
S\right)  =\sum_{u_{i}\in S}p_{i}$. Then $\rho\left(  \Sigma S\right)  $ is
again a pure state density matrix so it is obtained (up to sign) as the outer
product of the eigenvector $\left\vert s\right\rangle $ for the eigenvalue of
$1$ with its transpose where $\left\langle u_{i}|s\right\rangle =\sqrt
{\frac{p_{i}}{\Pr\left(  \Sigma S\right)  }}$ if $u_{i}\in S$, else $0$. Again
the new vector $\left\vert s\right\rangle $ for the pure state density matrix,
not the mixed state, is the vector of probability amplitudes and the Born rule
is $\Pr(u_{i}|\Sigma S)=\left\langle u_{i}|s\right\rangle ^{2}$.

These relationships continue to hold, \textit{mutatis mutandis}, in the case
of a pure state density matrix $\rho$ over the complex numbers. Again, there
is a vector of complex probability amplitudes $\left\vert \psi\right\rangle $,
which is the eigenvector for the eigenvalue of $1$, such that $\rho=\left\vert
\psi\right\rangle \left\langle \psi\right\vert $ where $\left\langle
\psi\right\vert $ is the conjugate transpose of $\left\vert \psi\right\rangle
$. Then the $i^{th}$ diagonal element of $\rho$ is obtained as the $i^{th}$
element of $\left\vert \psi\right\rangle $ times its complex conjugate which
is the $i^{th}$ element of the conjugate transpose $\left\langle
\psi\right\vert $, and that is the Born rule: $\Pr\left(  u_{i}|\rho\right)
=\left\vert \left\langle u_{i}|\psi\right\rangle \right\vert ^{2}$.

Rota was clear about his program for quantum probabilities: ``
I will lay my cards on the table: a revision of the notion of a sample space
is my ultimate concern.'' \cite[p. 57]{rota:fubini} In this
manner, we see how probability amplitudes and the Born rule arise from Rota's
program of redefining the sample space and its events. And it should not be a
surprise that the key non-classical notion in QM, superposition, was the way
to redefine them.

Perhaps a word is necessary about what constitutes an understanding of quantum
mechanics--as illustrated by this treatment of the Born rule. Given all the
mathematical machinery of QM in Hilbert space, there is a whole literature
\cite{ell:born-again} about ``deriving the Born
rule'' by showing that some result in QM would not hold
without the Born rule. There is even a completely mathematical result,
Gleason's Theorem \cite{gleason:born}, that is said to ``
explain'' the Born rule. But our contention is that these
types of ``explanations'' or
``derivations'' do not `do the job.' An
explanation \textit{that yields understanding} (as opposed to showing the
properties of a concept in a corpus of mathematics) needs to break it down
into simple intuitive terms that are `behind' the quantum phenomenon. That is
what we have attempted to do by showing how probability amplitudes and the
Born rule both result from adding the simple notion of a superposition set or
event to ordinary probability theory.

\section{From partition math to quantum math for states}\label{sec5}

\subsection{Using partition lattices to represent quantum states}\label{subsec8}

Let $U=\left\{  u_{1},...,u_{n}\right\}  $ be an orthonormal (ON) basis for
the $n$-dimensional complex Hilbert space. A quantum state is specified by a
normalized state vector $\left\vert \psi\right\rangle =\sum_{i=1}^{n}%
\alpha_{i}\left\vert u_{i}\right\rangle $ or by its pure state density matrix
$\rho=\left\vert \psi\right\rangle \left\langle \psi\right\vert $ with the
entries $\rho_{jk}=\alpha_{j}\alpha_{k}^{\ast}$ where $\alpha_{k}^{\ast}$ is
the conjugate transpose of $\alpha_{k}$. The diagonal elements $\rho
_{ii}=\alpha_{i}\alpha_{i}^{\ast}=q_{i}$ provide a probability distribution
$q=\left(  q_{1},...,q_{n}\right)  $ over basis $U$. Associated with such a
pure state density matrix over the complex numbers, there is a density matrix
$\rho^{\#}$ over the reals $%
\mathbb{R}
$ where $\rho_{jk}^{\#}=\sqrt{q_{j}q_{k}}=\sqrt{\alpha_{j}\alpha_{k}^{\ast
	}\alpha_{j}^{\ast}\alpha_{k}}$. A density matrix $\rho$ is pure iff
$\operatorname*{tr}\left[  \rho^{2}\right]  =1$ and the trace of a density
matrix squared is equal to the sum of the absolute squares of the entries in
the matrix \cite[p. 77]{fano:density} so that:

\begin{center}
	$1=\operatorname*{tr}\left(  \rho^{2}\right)  =\sum_{j,k}\left\vert \rho
	_{jk}\right\vert ^{2}=\sum_{j,k}\alpha_{j}\alpha_{k}^{\ast}\alpha_{j}^{\ast
	}\alpha_{k}=\sum_{j,k}q_{j}q_{k}=\operatorname*{tr}\left[  \left(  \rho
	^{\#}\right)  ^{2}\right]  .$
\end{center}

\noindent Thus $\rho^{\#}$is also a pure state density matrix and it carries
the same probability information as $\rho$. The \textit{support set} of
$\left\vert \psi\right\rangle =\sum_{i=1}^{n}\alpha_{i}\left\vert
u_{i}\right\rangle $ in the given basis $U$ is: $supp_{U}\left(  \left\vert
\psi\right\rangle \right)  =\left\{  u_{i}|\alpha_{i}\neq0\right\}  $. The
$q_{i}=\alpha_{i}\alpha_{i}^{\ast}$ for $u_{i}$ in the support set
$supp_{U}\left(  \left\vert \psi\right\rangle \right)  $ form a positive
probability distribution on the support set with $q_{i}=0$ outside the support
set. Previously, we started with the notion of a superposition event or state
$\Sigma S$ and developed that idea up to the density matrix $\rho\left(
\Sigma S\right)  $ over the reals. Now we have started with a pure state
density matrix $\rho$ over the complex numbers and constructed its counterpart
$\rho^{\#}$ over the reals. The two approaches meet since taking
$S=supp_{U}\left(  \left\vert \psi\right\rangle \right)  $, we see that:
$\rho\left(  \Sigma S\right)  =\rho^{\#}$ where $q_{i}=\frac{p_{i}}{\Pr\left(
	S\right)  }$--all of which confirms the matrix representation of a
superposition state in our previous treatment.

The \textit{support matrix} of a density matrix $\rho$ expressed in the
$U$-basis is the relation matrix $\operatorname*{Rel}\left(  \rho\right)  $
where $\operatorname*{Rel}\left(  \rho\right)  _{jk}=1$ if $\rho_{jk}\neq0$,
else $0$. In any (algebraic) field (as opposed to a ring) such as $
\mathbb{R}$ or $\mathbb{C}$, the product of two non-zero scalars is non-zero. Since any pure state
density matrix like $\rho$ can be obtained as the outer product $\left\vert
\psi\right\rangle \left\langle \psi\right\vert $ of a vector and its conjugate
transpose, we have the following result.

\begin{proposition}
	For any pure state density matrix $\rho=\left\vert \psi\right\rangle
	\left\langle \psi\right\vert $, $\operatorname*{Rel}\left(  \rho\right)
	=\operatorname*{Rel}\left(  supp_{U}\left(  \left\vert \psi\right\rangle
	\right)  \times supp_{U}\left(  \left\vert \psi\right\rangle \right)  \right)
	$.
\end{proposition}

\noindent In other words, for any pure state density matrix $\rho$ over the
complex numbers, the pattern of non-zero entries $\rho_{jk}$ in $\rho$ is
$S\times S$ for $S=supp_{U}\left(  \left\vert \psi\right\rangle \right)  $ as
in our previous notion of a superposition event $\Sigma S$.

Our goal in this section is to associate a set partition with every pure state
density matrix or every mixed state density matrix that results from a pure
state density matrix by a projective measurement modeled by the L\"{u}ders
mixture operation. The non-singleton blocks in the set partition will
correlate with superposition states and the singleton blocks with eigenstates.
All the non-singleton blocks $S$ associated with superpositions represent
superposition events $\Sigma S$ so we will henceforth drop the $\Sigma$
notion; all non-singleton blocks represent superpositions.

We already have the answer starting with a pure state density matrix
$\rho=\left\vert \psi\right\rangle \left\langle \psi\right\vert $. Cutting
down the set $U$ to $S=supp_{U}\left(  \left\vert \psi\right\rangle \right)  $
with the probability distribution $q_{i}=\frac{p_{i}}{\Pr\left(  S\right)  }$
for $u_{i}\in S$, the associated set partition is the indiscrete partition
$\mathbf{0}_{S}$. Since we can always cut down the universe set $U$ with a
probability distribution $p$ to a subset $S$ with positive probabilities
$q_{i}=\frac{p_{i}}{\Pr\left(  S\right)  }$, we will henceforth just assume
that the distribution $p$ on $U$ is strictly positive.

Given a set partition $\pi=\left\{  B_{1},...,B_{m}\right\}  $ on $U$ with
positive point probabilities $p$, we already have a pure state density matrix
$\rho\left(  B_{j}\right)  $ [where we write $\rho\left(  B_{j}\right)  $
instead of $\rho\left(  \Sigma B_{j}\right)  $] associated with each block.
Then we define a mixed state density matrix $\rho\left(  \pi\right)  $
associated with $\pi$ as the probabilistic sum of the pure state matrices for
the blocks:

\begin{center}
	$\rho\left(  \pi\right)  =\sum_{j=1}^{m}\Pr\left(  B_{j}\right)  \rho\left(
	B_{j}\right)  $.
\end{center}

Density matrices can be presented as the probabilistic sum of pure state
density matrices in numerous ways \cite[p. 105]{nc:qcqi}. The matrix
$\rho\left(  \pi\right)  $ has the special property that it is the
probabilistic sum of pure state matrices of disjoint supports $B_{j}$. The
quantum states that we wish to model at the level of set partitions are the
states that can be obtained by a projective measurement of a pure state. This
is because the mixed state density matrices $\hat{\rho}$ that are the results
of projective measurements of pure state density matrices $\rho$ have the
special property that they are the probabilistic sums of pure state density
matrices with \textit{disjoint} supports.

We take $U$ to be an orthonormal (ON) basis of eigenvectors of an observable
(Hermitian or self-adjoint) operator $F:V\rightarrow V$. The eigenvalues are
all real so there is an \textit{eigenvalue function} $f:U\rightarrow
\mathbb{R}$ assigning each eigenvector its eigenvalue. The eigenvalue function defines
the inverse-image set partition $f^{-1}=\left\{  f^{-1}\left(  r\right)
\right\}  _{r\in f\left(  U\right)  }$. The eigenspace of $F$ associated with
$r\in f\left(  U\right)  $ is the subspace $\left[  f^{-1}\left(  r\right)
\right]  $ of $V$ generated by the eigenvectors in that block of $f^{-1}$. Let
$P_{r}:V\rightarrow V$ be the projection operator to that eigenspace $\left[
f^{-1}\left(  r\right)  \right]  $. Given a pure state density matrix $\rho$
represented in the $U$-basis, let $p_{i}$ be the diagonal probability entry in
$\rho$ associated with $u_{i}\in U$. Then the L\"{u}ders mixture operation
\cite[p. 279]{auletta:qm} gives the mixed state $\hat{\rho}$ resulting from
the projective $F$-measurement of $\rho$:

\begin{center}
	$\hat{\rho}=\sum_{r\in f\left(  U\right)  }P_{r}\rho P_{r}=\sum_{r\in f\left(
		U\right)  }\Pr\left(  f^{-1}\left(  r\right)  \right)  \frac{P_{r}\rho P_{r}%
	}{\Pr\left(  f^{-1}\left(  r\right)  \right)  }$
\end{center}

\noindent where $\Pr\left(  f^{-1}\left(  r\right)  \right)  =\sum_{u_{i}\in
	f^{-1}\left(  r\right)  }p_{i}$.

\begin{proposition}
	The result $\hat{\rho}=\sum_{r\in f\left(  U\right)  }P_{r}\rho P_{r}%
	=\sum_{r\in f\left(  U\right)  }\Pr\left(  f^{-1}\left(  r\right)  \right)
	\frac{P_{r}\rho P_{r}}{\Pr\left(  f^{-1}\left(  r\right)  \right)  }$ of the
	projective $F$-measurement of a pure state density matrix $\rho$ in QM can be
	represented as the probability sum of pure state density matrices which are
	the outer products of vectors with \textbf{disjoint} supports.
\end{proposition}

\textbf{Proof}: It needs to be shown that$\frac{P_{r}\rho P_{r}}{\Pr\left(
	f^{-1}\left(  r\right)  \right)  }$ is a pure state density matrix. The
general form is for some subset $S\subseteq U$, $\frac{1}{\Pr\left(  S\right)
}P_{S}\rho P_{S}$ where $P_{S}=\sum_{u_{i}\in S}\left\vert u_{i}\right\rangle
\left\langle u_{i}\right\vert $ which is the diagonal matrix with entries
$\chi_{S}\left(  u_{i}\right)  $. The point probabilities on $U$ are
$p_{i}=\rho_{ii}$ so $\Pr\left(  S\right)  =\sum_{u_{i}\in S}p_{i}$. Since
$\rho$ is pure state density matrix, there is a normalized vector $\left\vert
s\right\rangle $ such that $\rho=\left\vert s\right\rangle \left\langle
s\right\vert $. Then $P_{S}\rho P_{S}=P_{S}\left\vert s\right\rangle
\left\langle s\right\vert P_{S}$ so let $\left\vert s^{\prime}\right\rangle
=\frac{1}{\sqrt{\Pr(S)}}P_{S}\left\vert s\right\rangle $ so that $\left\vert
s^{\prime}\right\rangle \left\langle s^{\prime}\right\vert =\frac{1}%
{\Pr\left(  S\right)  }P_{S}\rho P_{S}$. To show that it is a density matrix,

\begin{center}
	$\operatorname*{tr}\left[  \left\vert s^{\prime}\right\rangle \left\langle
	s^{\prime}\right\vert \right]  =\frac{1}{\Pr\left(  S\right)  }%
	\operatorname*{tr}\left[  P_{S}\rho P_{S}\right]  =\frac{1}{\Pr\left(
		S\right)  }\operatorname*{tr}\left[  \rho P_{S}P_{S}\right]  =\frac{1}%
	{\Pr\left(  S\right)  }\operatorname*{tr}\left[  \rho P_{S}\right]  =\frac
	{1}{\Pr\left(  S\right)  }\sum_{u_{i}\in S}\rho_{ii}=1$.
\end{center}

\noindent Hence $\left\vert s^{\prime}\right\rangle \left\langle s^{\prime
}\right\vert $ is a pure state density matrix so taking $S=f^{-1}\left(
r\right)  $, so is $\frac{P_{r}\rho P_{r}}{\Pr\left(  f^{-1}\left(  r\right)
	\right)  }$. The support set of $\left\vert s^{\prime}\right\rangle $ for
$S=f^{-1}\left(  r\right)  $ is $f^{-1}\left(  r\right)  $ and for
$r,r^{\prime}\in f\left(  U\right)  $ with $r\neq r^{\prime}$, $f^{-1}\left(
r\right)  \cap f^{-1}\left(  r^{\prime}\right)  =\emptyset$ so the outer
product vectors have disjoint supports. The disjointness results from the
orthogonality of the eigenspaces of the observable. $\square$

Our goal in this section is to associate the lattice of partitions $\Pi\left(
U\right)  $ with a state vector $\left\vert \psi\right\rangle $ with the
support set $U$. The bottom is the indiscrete partition $\mathbf{0}%
_{U}=\left\{  U\right\}  $ and the top is discrete partition $\mathbf{1}%
_{U}=\left\{  \left\{  u_{i}\right\}  \right\}  _{u_{i}\in U}$. The indiscrete
partition $\mathbf{0}_{U}$ is associated with the pure state density matrix
$\rho=\left\vert \psi\right\rangle \left\langle \psi\right\vert $. The
non-singleton blocks $B_{j}$ in the partitions of $\Pi\left(  U\right)  $
represent superpositions of the elements in the block which we previously
denoted as $\Sigma B_{j}$. The singleton blocks represent eigenstates. The
discrete partition $\mathbf{1}_{U}$ is all singletons so it represents the
completely decomposed mixed state which we previously associated with the
classical discrete event $U$. All the partitions in between represent the
mixed states that could result from projective measurements of $\rho
=\left\vert \psi\right\rangle \left\langle \psi\right\vert $.

The view of reality implicit in classical physics was that reality was
definite `all the way down.' This was embodied in Leibniz's Principle of
Identity of Indistinguishables (PII). If two entities were actually different,
then by going `down far enough', there would be some predicate that applied to
one and not the other. Otherwise, they are indistinguishable and Leibniz's
principle says they must be identical. That is the principle of classicality.

The discrete partition $\mathbf{1}_{U}$ representing completely decomposed
mixed quantum states satisfies the corresponding partition version of the PII.
That is, if $\left(  u,u^{\prime}\right)  \in\operatorname*{indit}\left(
\mathbf{1}_{U}\right)  $ are indistinct according to the discrete partition,
then they are identical $u=u^{\prime}$. This is trivial mathematically since
the discrete partition makes all the distinctions that can be made so the only
indistinctions left are the self-pairs $\left(  u_{i},u_{i}\right)  $. In
short, there is nothing quantum about the state represented by the discrete
partition; it is like ``the statistical mixture describing the
state of a classical dice before the outcome of the throw''%
\ \cite[p. 176]{auletta:qm}. This partition version of the PII holds only for
the discrete partition; all the other partitions in the lattice contain at
least one non-singleton block which represents a superposition so, aside from
the `pure state' indiscrete partition, all the in-between partitions represent
non-classical mixed states.

The important thing about this partition lattice $\Pi\left(  U\right)  $ is
that represents in a skeletal or essential way a pure superposition state and
all the other states that can result from measuring it. The lattice divides
into the classical part, only $\mathbf{1}_{U}$, and the quantum part beneath
it in the lattice so the lattice can be thought of as an iceberg (the
classical part above the water and the quantum part below the water). Figure 4
gives this lattice for $U=\left\{  a,b,c,d\right\}  $ (where the innermost
curly brackets are deleted so the partition $\left\{  \left\{  a,b,c\right\}
,\left\{  d\right\}  \right\}  $ is just $\left\{  abc,d\right\}  $). Since
the elements in the blocks represent states, the whole lattice represents the
possible states of a quantum particle in a four-dimensional Hilbert space.

\begin{figure}[h]
	\centering
	\includegraphics[width=0.9\linewidth]{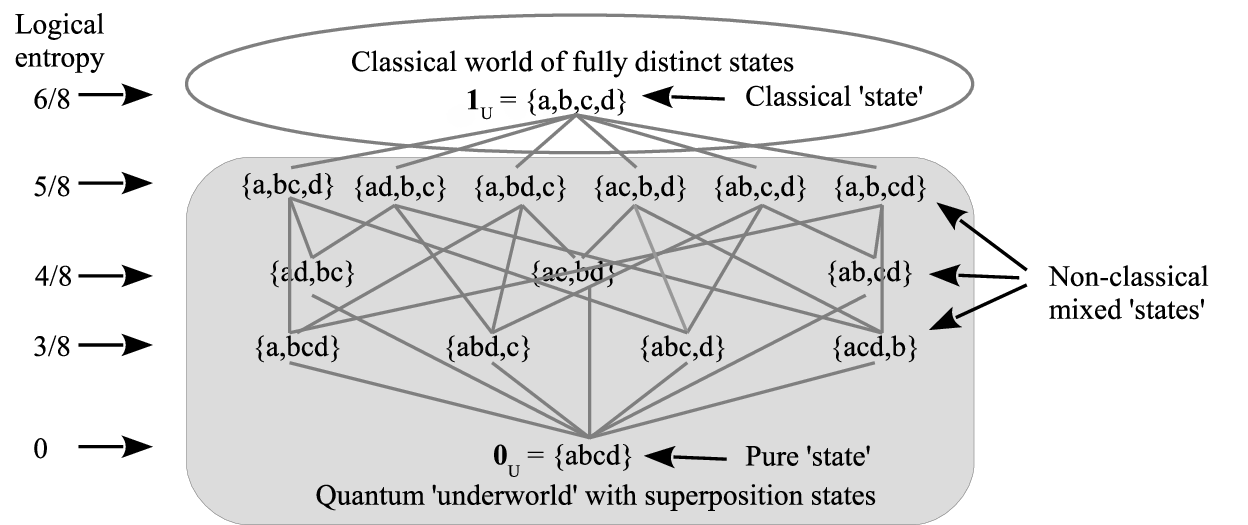}
	\caption{The lattice of partitions representing the possible states of a
		quantum particle with four eigenstates}
	\label{fig:4-lattice-iceberg}
\end{figure}

\noindent The logical entropies for the partitions assume equiprobable points.
The idea is to picture the essentials of the possible states of a quantum
system. In picturing the ``essentials'' using
partitions of support sets, certain information is ignored, e.g., whether a
non-zero coefficient is positive or negative, or a complex number or its
conjugate, and so forth. For a simple example, consider the Bell state of two
particles with opposite spins in a certain direction before and after a
measurement in Figure 5 where $U=\left\{  1\downarrow\otimes2\uparrow
,1\uparrow\otimes2\downarrow\right\}  $ so the only two partitions are the
indiscrete and discrete partitions.

\begin{figure}[h]
	\centering
	\includegraphics[width=0.7\linewidth]{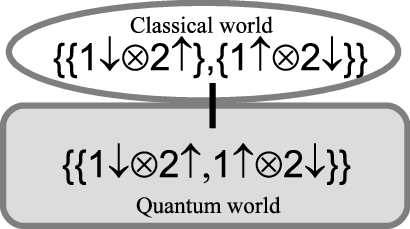}
	\caption{Iceberg diagram of Bell state with opposite spins}
	\label{fig:bell-state}
\end{figure}

\subsection{Simple application to the double-slit experiment}\label{subsec9}

The two parts of the iceberg (partition lattice) diagrams allow us to
illustrate in a simplified manner what happens in the two cases (detection or
no detection at the slits) of the two-slit experiment. Feynman said that this
experiment illustrates ``the only mystery'' \cite[Sec. 1.1]{feynman:v3mill-ed} in quantum mechanics. That mystery is how
to answer the question about the particle in the case of no detection at the
slits: ``which hole does it go through?''.
\cite[Sec. 2.16]{feynman:v3mill-ed} As Feynman repeatedly pointed out,
``We must conclude that when both holes are open it is `not
true' that the particle goes through one hole or the other.''%
\ \cite[p. 536]{feynman: prob-in-qm} This is because if the particle goes
through one of the slits, then there would be no interference phenomena.

\begin{quotation}
	\noindent Feynman's approach is based on the contrast between processes that
	are \textit{distinguishable} within a given physical context and those that
	are \textit{indistinguishable} within that context. A process is
	distinguishable if some record of whether or not it has been realized results
	from the process in question; if no record results, the process is
	indistinguishable from alternative processes leading to the same end result.
	In my terminology, a registration of the realization of a process must exist
	for it to be a distinguishable alternative. In the two-slit experiment, for
	example, passage through one slit or the other is only a distinguishable
	alternative if a counter is placed behind one of the slits; without such a
	counter, these are indistinguishable alternatives. Classical probability rules
	apply to distinguishable processes. Nonclassical probability amplitude rules
	apply to indistinguishable processes. \cite[p. 314]{stachel:qlogic}
\end{quotation}

\noindent What Feynman called ``
distinguishability,'' Anton Zeilinger calls ``
information.''

\begin{quotation}
	\noindent In other words, the superposition of amplitudes ... is only valid if
	there is no way to know, even in principle, which path the particle took. It
	is important to realize that this does not imply that an observer actually
	takes note of what happens. It is sufficient to destroy the interference
	pattern, if the path information is accessible in principle from the
	experiment or even if it is dispersed in the environment and beyond any
	technical possibility to be recovered, but in principle still
	``out there.'' The absence of any such
	information is the essential criterion for quantum interference to appear.
	\cite[p. 484]{zeilinger:1999}
\end{quotation}

It might be noted that \v{C}aslav Bruckner and Anton Zeilinger have also
recommended the formula for logical entropy to measure quantum information
\cite[p. 332]{bruk-zeil:2003} and as Charles Bennett (one of the founders of
quantum information has argued): ``So information really is a
very useful abstraction. It is the notion of distinguishability abstracted
away from what we are distinguishing, or from the carrier of information ...
.'' \cite[p. 155]{bennett:qinfo}

For most of a century, most quantum theorists have `explained' the
no-detection case of the two-slit experiment by resorting to the magic of
`wave-particle complementarity.' With no detection at the slits, the particle
suddenly turns into an ontic wave, goes through both slits, and then shows the
interference phenomena like in the (misleading) classroom ripple-tank version
of the experiment, and then turns back into a particle when it hits the
detection wall. The ice-berg/partition-lattice diagram shows how to understand
what happens without resorting to Jekyll-and-Hyde version of a quantum particle.

There are two fundamentally different processes in quantum mechanics which von
Neumann \cite{vonn:mfqm} called ``Type I''%
\ processes (state reductions from indefinite to less-indefinite or definite
states) and ``Type II'' processes (unitary
evolution). These two types of processes have simple representations in the
lattice diagrams.

\begin{itemize}
	\item Type I: Since the ordering of partitions is refinement (inclusion
	relation between ditsets), a movement upward in the diagram represents a
	making of more distinctions so that indefinite becomes less-indefinite if not
	definite as a type I process.This is how definite classical states emerge from the quantum underworld of indefiniteness by the making of distinctions or distinguishings as in the Feynman rules.
	
	\item Type II: And conversely, any non-upward (horizontal or downward)
	movement in a lattice diagram represents a type II process of unitary evolution.
\end{itemize}

This is illustrated in Figure 6 where $\left\{  a,c\right\}  $ represents the
superposition state of ``Going through slit 1 + Going through
slit 2''. Going through one slit $\left\{  a\right\}  $ or the
other $\left\{  c\right\}  $ is a classical event in the lattice diagram but
in the case of no detection at the slits, there is no distinguishing between
the superposed states so it does not rise to the classical level of $\left\{
a\right\}  $ or $\left\{  c\right\}  $ and then evolve (the crossed-out dotted
lines in Figure 6). Instead, it evolves `horizontally' (the solid line) to
another superposition state represented in Figure 6 as $\left\{  a^{\prime
},c^{\prime}\right\}  $. That type II evolution is in the quantum part of
reality (the underwater part of the iceberg) which is why it cannot be
explained in classical (above-water) terms. By understanding why superposition
states will continue to evolve as superposition states when the superposed
alternatives are not distinguished, one avoids having to resort to the magic
of `wave-particle duality' to `explain' what happens (see the treatment of the
double-slit experiment below in the pedagogical model of QM/Sets, quantum
mechanics over sets).

\begin{figure}[h]
	\centering
	\includegraphics[width=0.7\linewidth]{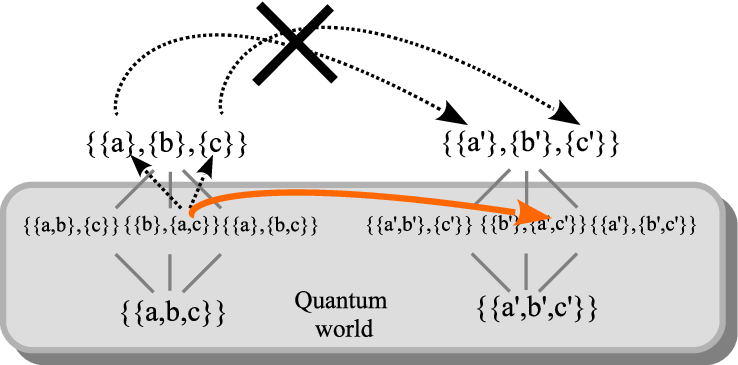}
	\caption{The type II evolution of a superposition with no distinguishing
		between the alternative paths of going one slit or the other}
	\label{fig:case2-double-slit}
\end{figure}

The cost of this explanation is giving up on trying to explain everything in
classical terms (or resorting to magic). The alternative is understanding that
there is, at the quantum level, a world of indefiniteness, and that the
classical definite world only emerges as distinctions transform indefinite
states into definite states. This suggests a non-classical interpretation of superposition.

\section{How to define quantum superposition}\label{sec6}

Partitions (or equivalence relations) are the abstract mathematical concept to
model identity versus difference, equivalence versus inequivalence, or
indefiniteness versus definiteness. The superposition principle is the key
non-classical notion in QM but superposition is usually misinterpreted as
being like the superposition of classical waves like water or electomagnetic
waves where definite waves are added to definite waves to give an equally
definite wave as illustrated in Figure 7.

\begin{figure}[h]
	\centering
	\includegraphics[width=0.7\linewidth]{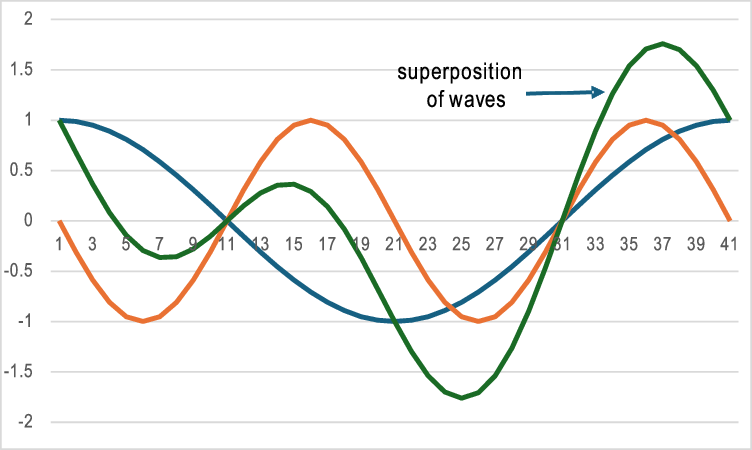}
	\caption{Classical wave superposition: definite + definite = definite}
	\label{fig:wavesuperposition}
\end{figure}

Quantum superposition is different. The superposition of two quantum definite-
or eigen-states is \textit{indefinite} on the differences between them as
illustrated in the simple example of Figure 8. On the real $XY$-plane, the
point $\left(  1,0\right)  $ is definite in terms of $X$-ness and the point
$\left(  0,1\right)  $ is definite in terms of $Y$-ness, but their
superposition $\left(  1,1\right)  $ is indefinite between pure $X$-ness and
pure $Y$-ness. This reinterpretation of superposition is a key point to our later
treatment of state reduction.

\begin{figure}[h]
	\centering
	\includegraphics[width=0.7\linewidth]{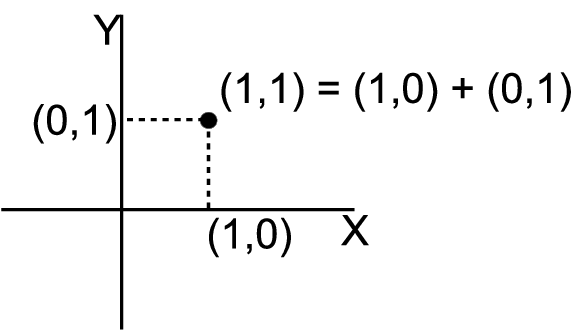}
	\caption{Simple illustration of definite $X$-ness + definite $Y$-ness =
		indefinite $\left(  X,Y\right)  $-ness}
	\label{fig:xy-siuperposiiton}
\end{figure}

Quantum mechanics cannot be modeled solely in terms of classical fully
definite reality--which is modeled here by the discrete partition that
represents classicality. Reality is represented by the whole `iceberg.' There
is the quantum `under-world' of indefiniteness; the fully definite classical
world is constantly emerging from the quantum level of superposed indefinite
reality by interactions that make distinctions called ``state
reductions'' (also called ``
measurements'' although that word unnecessarily connotes
human involvement). Any non-philosophical quantum theorist who acknowledges
that a quantum (superposition) state in an observable does not have a definite
value prior to a measurement also automatically acknowledges the reality of
this quantum underworld of indefiniteness.

Quantum theorists such as Werner Heisenberg and Abner Shimony (and many
others) have referred to the quantum world as existing in potentiality so that
reality consists of the classical definite world and the quantum world of
indefiniteness (where definiteness is only a potentiality). In the whole
`demolition derby' of quantum interpretations, Shimony suggested this as the
``Literal Interpretation.''

\begin{quotation}
	\noindent Heisenberg (\cite[p. 53]{heisen:phy-phil}) used the term
	``potentiality'' to characterize a property
	which is objectively indefinite, whose value when actualized is a matter of
	objective chance, and which is assigned a definite probability by an algorithm
	presupposing a definite mathematical structure of states and properties.
	Potentiality is a modality that is somehow intermediate between actuality and
	mere logical possibility. That properties can have this modality, and that
	states of physical systems are characterized partially by the potentialities
	they determine and not just by the catalogue of properties to which they
	assign definite values, are profound discoveries about the world, rather than
	about human knowledge. It is fair to say, in view of my discussion above of
	metaphysics, that these statements about quantum mechanical potentiality are
	metaphysical propositions suggested by the formalism of quantum mechanics.
	These statements, together with the theses about potentiality, may
	collectively be called ``the Literal
	Interpretation'' of quantum mechanics. This is the
	interpretation resulting from taking the formalism of quantum mechanics
	literally, as giving a representation of physical properties themselves,
	rather than of human knowledge of them, and by taking this representation to
	be complete. \cite[pp. 6-7]{shim:vienna-yrbk}
\end{quotation}

\noindent Henry Margenau \cite{margenau:latency} and R. I. G. Hughes
\cite{hughes:structure} preferred the word ``
latency'' as being less laden with Aristotelean overtones of
``potentia'' or ``
potentiality'' but the idea was the same.

\section{Linearization: From partition math to QM math}\label{sec7}

\subsection{The Yoga of Linearization}\label{subsec10}

Linear algebra was a powerful tool in the Rota school of combinatorial theory.
In the mathematical folklore, there is a well-known connection between set
concepts and the corresponding or analogous linear algebra concept. By taking
the linear algebra concept to be a Hilbert space of quantum mechanics, one
obtains the quantum analog of set-based concepts. ``The
quantum analog of a random variable is called an observable.''%
\ \cite[p. 81]{rota:fubini} These analogies, developed systematically, are
what Rota might call a ``yoga'' \cite[p.
251]{rota:indiscrete}, the yoga of linearization, e.g., as developed by Steven
Roman \cite[pp. 355-361]{roman:adv-lin-alg}, one of Rota's students and
collaborators. These analogies are extensively developed in Rota's enumerative
combinatorics as ``q-analogs'' where the
vector space is over a finite field of order $q$ \cite{goldman-rota:IV}, e.g.,
``Direct-sum decompositions are a $q$-analog of partitions of
a finite set.'' \cite[p. 764]{bender-goldman:genfcns}

Here is one simple way to develop the:
\begin{center}%
	\begin{tabular}
		[c]{|c|}\hline
		Yoga of Linearization\\\hline\hline
		Given a basis set of a vector space, apply the set concept to the basis set\\
		and then what is linearly generated is the corresponding or analog vector
		space concept.\\\hline
	\end{tabular}
	
\end{center}

Our task is to develop the dictionary of analogs when the set concepts are the
set mathematics related to partitions, i.e., `partition math,' and the vector
spaces are the Hilbert spaces of quantum mechanics, i.e., from the set math of
partitions to their QM analogs. We stick to the finite-dimensional case since
our goal is conceptual clarification rather mathematical generality and,
according to Hermann Weyl, ``no essential features of quantum
mechanics are lost by using the finite-dimensional model.'' \cite[p. 257]{weyl:phil} Our goal is to show firstly that the linearization
of the math of partitions gives the essential math of QM, and then to develop
the implications for understanding the reality that is modeled by QM. Once the
dictionary of set/vector-space analog concepts is developed, the set concepts
can be represented in a vector space over $\mathbb{Z}_{2}$. In that manner, we develop a much simplified pedagogical model of QM
over $\mathbb{C}$ as QM over $\mathbb{Z}_{2}$ or QM/Sets \cite{ell:simplified}.

\subsection{Developing the partition-math to QM-math dictionary for
	observables}\label{subsec11}

We need to develop the partition math versions of QM states and QM
observables. Previously, it was shown how quantum states could be represented
in a skeletal form by partitions, e.g., in the iceberg diagram. Now we turn to
quantum observables.

Let $U=\left\{  u_{1},...,u_{n}\right\}  $ be an orthonormal (ON) basis for
the $n$-dimensional complex Hilbert space $V=\mathbb{C}^{n}$ (where we will alternatively take $u_{i}$ as an element of $U$ as a set
or as a vector, also denoted as $\left\vert u_{i}\right\rangle $, in the
Hilbert space). A subset $S\subseteq U$ generates the subspace $\left[
S\right]  \subseteq V$ or as Rota put it: ``Since von Neumann
it has been agreed that closed subspaces of a Hilbert space are the quantum
mechanical analogs of probabilistic events.'' \cite[p.
154]{rota:indiscrete} The cardinality $\left\vert S\right\vert $ of $S$
correlates with the dimension $\dim\left(  \left[  S\right]  \right)  $ of
$\left[  S\right]  $. Rota's statement that the ``quantum
analog of a random variable is called an observable'' is not
quite right since observables have no probability information in them; the
probabilities are supplied by the quantum state. If we delete the probability
information from the domain $U$ of a numerical attribute $f:U\rightarrow \mathbb{R}$, then the quantum analog is the observable or Hermitian operator $\hat{f}$
defined by $\hat{f}u_{i}=f\left(  u_{i}\right)  u_{i}$ on the basis set $U$
and then extended linearly to $\hat{f}:V\rightarrow V$. This correlation
between $f$ and $\hat{f}$ is reversible. Given an observable $\hat
{f}:V\rightarrow V$ , it has an orthonormal basis $U$ of eigenvectors and thus
an \textit{eigenvalue function} $f:U\rightarrow \mathbb{R}$ such that $\hat{f}u_{i}=f\left(  u_{i}\right)  u_{i}$. If for $S\subseteq
U$, we interpret the equation $f\upharpoonright S=rS$ to mean that $f$
restricted to $S$ has the constant value $r$, then that equation is the set
version of the usual eigenvalue-eigenvector equation: $\hat{f}u_{i}=ru_{i}$
which means that the set-version of an eigenvector is a constant set of a
numerical attribute and the constant value $r$ is the set-version of the
eigenvalue. A numerical attribute $f:U\rightarrow \mathbb{R}$ defines the inverse-image partition $f^{-1}=\left\{  f^{-1}\left(  r\right)
\right\}  _{r\in f\left(  U\right)  }$ and each block $f^{-1}\left(  r\right)
$ generates the subspace $\left[  f^{-1}\left(  r\right)  \right]  $, and
those subspaces, the eigenspaces of $\hat{f}$, make up a direct-sum
decomposition (DSD) of $V$. Thus the vector space version of a partition on
$U$ is a DSD of the vector space generated by $U$. These dictionary entries
are summarized in Table 4.

\begin{center}%
	\begin{tabular}
		[c]{|c|c|}\hline
		Set math & QM analogs\\\hline\hline
		Subset $S\subseteq U=\left\{  u_{1},...,u_{n}\right\}  $ & Subspace $\left[
		S\right]  \subseteq V$\\\hline
		Cardinality $\left\vert S\right\vert $ & Dimension $\dim\left(  \left[
		S\right]  \right)  $\\\hline
		Numerical attribute $f:U\rightarrow	\mathbb{R}$ & Hermitian op. $\hat{f}u_{i}=f\left(  u_{i}\right)  u_{i}$\\\hline
		Constant set $\ f\upharpoonright S=rS$, $r\in f\left(  U\right)  $ &
		Eigenvector $\hat{f}v=rv$\\\hline
		Value $r$ of $f$ & Eigenvalue $r$ of $\hat{f}$\\\hline
		Set of constant $r$-sets $\wp\left(  f^{-1}\left(  r\right)  \right)  $ &
		Eigenspace of $r$, $\left[  f^{-1}\left(  r\right)  \right]  $\\\hline
		Partition: $\left\{  f^{-1}\left(  r\right)  \right\}  _{r\in f\left(
			U\right)  }$ & Direct-Sum Decomposition: $\left\{  \left[  f^{-1}\left(
		r\right)  \right]  \right\}  _{r\in f\left(  U\right)  }$\\
		$U=\uplus_{r\in f\left(  U\right)  }f^{-1}\left(  r\right)  $ & $V=\oplus
		_{r\in f\left(  U\right)  }\left[  f^{-1}\left(  r\right)  \right]  $\\\hline
	\end{tabular}

	Table 4: Partial dictionary or correlation table between partition math and
	their QM (Hilbert space) analogs
\end{center}

A special case of a numerical attribute is a characteristic function for a
subset $S\subseteq U$, $\chi_{S}:U\rightarrow\left\{  0,1\right\}  $, and the
operator defined by a characteristic function is a projection operator.
Observables have a spectral resolution as the sum of the eigenvalues times the
projections to the eigenspaces of those eigenvalues. The set version is a
`spectral resolution' of a numerical attribute $f:U\rightarrow \mathbb{R}$ where characteristic functions have the role of their quantum analogs,
projection operators. In a vector space, there are, of course, many different
basis sets the same cardinality as $U$. One of the most striking correlations
comes from the fact that different numerical attributes
$f,g,...,h:U\rightarrow \mathbb{R}$ all defined on the same $U$ will define observables $\hat{f},\hat
{g},...,\hat{h}$ that are commuting (since $U$ provides a set of simultaneous
eigenvectors for all of them). On the set side, we can take the join of the
partitions $f^{-1},g^{-1},...,h^{-1}$ and if that join equals the discrete
partition $\mathbf{1}_{U}$ with blocks of cardinality one, then each $u_{i}\in
U$ is uniquely characterized by the ordered tuple of its `eigenvalues'
$\left(  f\left(  u_{i}\right)  ,g\left(  u_{i}\right)  ,...,h\left(
u_{i}\right)  \right)  $. If the numerical attributes $f,g,...,h$ are defined
on the same set, then they are said to be \textit{compatible} and if the join
of their inverse-image partitions is the discrete partition $\mathbf{1}_{U}$,
then they are said to be a \textit{complete }set of compatible attributes or a CSCA.

That partition math translates directly to Hilbert space math to give Dirac's
Complete Set of Commuting Observables or CSCO \cite[p. 57]{dirac:principles}.
Just as the partition join is formed by the non-empty intersection of the
blocks of the two partitions, so the join of the direct-sum decompositions
(i.e., vector space partition) of commuting observables consists of the
subspaces that are the non-zero-space intersections of the subspaces of the
DSDs. When the join of the DSDs for the commuting observables $\hat{f},\hat
{g},...,\hat{h}$ has subspaces all of dimension one (i.e., rays), then the
simultaneous eigenvectors are each uniquely characterized by the ordered tuple
of the eigenvalues. And if we extend the `linearly generated' to `bilinearly
generated', then the Cartesian product of basis sets bilinearly generates the
tensor product of the vector spaces. These correlations between partition math
and Hilbert space math are summarized in Table 5.

\begin{center}%
	\begin{tabular}
		[c]{|c|c|}\hline
		Set math & QM (Hilbert space) analogs\\\hline\hline
		$\chi$-function $\chi_{f^{-1}\left(  r\right)  }:U\rightarrow\left\{
		0,1\right\}  $ & {\small Projection op.} $P_{r}u_{i}=\chi_{f^{-1}\left(
			r\right)  }\left(  u_{i}\right)  u_{i}$\\\hline
		{\small Spectral decomposition} $f=\sum_{r\in f\left(  U\right)  }%
		r\chi_{f^{-1}\left(  r\right)  }$ & $\hat{f}=\sum_{r\in f\left(  U\right)
		}rP_{r}$\\\hline
		Numerical attributes $f,g,...,h:U\rightarrow \mathbb{R}	$ & Commuting operators $\hat{f},\hat{g},...,\hat{h}$\\\hline
		$U=$ same domain of $f,g,...,h$ & $U=$ {\small Simultaneous eigenvectors}
		$\hat{f},\hat{g},...,\hat{h}$\\\hline
		Join $f^{-1}\vee g^{-1}$ $=\left\{  f^{-1}\left(  r\right)  \cap g^{-1}\left(
		s\right)  \neq\emptyset\right\}  $ & $\left\{  \left[  f^{-1}\left(  r\right)
		\right]  \cap\left[  g^{-1}\left(  s\right)  \right]  \neq\left\{  0\right\}
		\right\}  $\\\hline
		$f^{-1}\vee g^{-1}\vee...\vee h^{-1}=\left\{  \left\{  u_{i}\right\}
		\right\}  _{u_{i}\in U}$ & $\hat{f}\vee\hat{g}\vee...\vee\hat{h}=$ DSD of
		rays\\
		$u_{i}\leftrightarrow\left(  f\left(  u_{i}\right)  ,g\left(  u_{i}\right)
		,...,h\left(  u_{i}\right)  \right)  $ & $u_{i}\leftrightarrow\left(  f\left(
		u_{i}\right)  ,g\left(  u_{i}\right)  ,...,h\left(  u_{i}\right)  \right)  $\\
		Set version CSCA & Dirac's CSCO\\\hline
		$U\times U=\left\{  \left(  u,u^{\prime}\right)  |u,u^{\prime}\in U\right\}  $
		& $V\otimes V=\left[  \left\{  u\otimes u^{\prime}|u,u^{\prime}\in U\right\}
		\right]  $\\\hline
	\end{tabular}

	Table 5: More partition math and their QM (Hilbert space) analogs
\end{center}

\section{The pedagogical model of QM/Sets}\label{sec8}

\subsection{The basics}\label{subsec12}

The idea is to collect together the set concepts correlated with the Hilbert
space analogs of full QM over $\mathbb{C}$, and then to represent the set concepts in vector spaces over $\mathbb{Z}_{2}=\left\{  0,1\right\}  $ as quantum mechanics over sets or QM/Sets
\cite{ell:simplified}. Multiplication in that field is as usual and addition
mod $2$ is only different since $1+1=0$. Column vectors in $\mathbb{Z}_{2}^{n}$ represent a subset $S\subseteq U$ with the $0,1$-entries of the
characteristic function $\chi_{S}:U\rightarrow\left\{  0,1\right\}  $. It is
more visual if we take the elements of $U$ as distinct elements like
$U=\left\{  a,b,c\right\}  $ and then a subset $S\subseteq U$ is just a set of
those elements, e.g., $S=\left\{  b,c\right\}  $. Then the vector space can be
seen as the powerset $\wp\left(  U\right)  $ rather than $\mathbb{Z}_{2}^{n}$. Addition mod $2$ is then represented by the symmetric difference
operation on subsets, e.g.,

\begin{center}
	$\left\{  a,b\right\}  +\left\{  b,c\right\}  =\left(  \left\{  a,c\right\}
	-\left\{  b,c\right\}  \right)  \cup\left(  \left\{  b,c\right\}  -\left\{
	a,b\right\}  \right)  =\left\{  a\right\}  \cup\left\{  c\right\}  =\left\{
	a,c\right\}  $.
\end{center}

One advantage of representing the set concepts as vectors is that there is
then a number of different basis sets in the vector space and that is
important in constructing analogs of Hilbert space concepts. Table 6 gives
some basis sets where the second one is for $\wp\left(  U^{\prime}\right)  $
where $U^{\prime}=\left\{  a^{\prime},b^{\prime},c^{\prime}\right\}  $ and the
third one for $U^{\prime\prime}=\left\{  a^{\prime\prime},b^{\prime\prime
},c^{\prime\prime}\right\}  $.

\begin{center}%
	\begin{tabular}
		[c]{|c|c|c|}\hline
		$U=\left\{  a,b,c\right\}  $ & $U^{\prime}=\left\{  a^{\prime},b^{\prime
		},c^{\prime}\right\}  $ & $U^{\prime\prime}=\left\{  a^{\prime\prime
		},b^{\prime\prime},c^{\prime\prime}\right\}  $\\\hline\hline
		$\left\{  a,b,c\right\}  $ & $\left\{  b^{\prime}\right\}  $ & $\left\{
		a^{\prime\prime},b^{\prime\prime},c^{\prime\prime}\right\}  $\\\hline
		$\left\{  a,b\right\}  $ & $\left\{  a^{\prime}\right\}  $ & $\left\{
		b^{\prime\prime}\right\}  $\\\hline
		$\left\{  b,c\right\}  $ & $\left\{  c^{\prime}\right\}  $ & $\left\{
		b^{\prime\prime},c^{\prime\prime}\right\}  $\\\hline
		$\left\{  a,c\right\}  $ & $\left\{  a^{\prime},c^{\prime}\right\}  $ &
		$\left\{  c^{\prime\prime}\right\}  $\\\hline
		$\left\{  a\right\}  $ & $\left\{  b^{\prime},c^{\prime}\right\}  $ &
		$\left\{  a^{\prime\prime}\right\}  $\\\hline
		$\left\{  b\right\}  $ & $\left\{  a^{\prime},b^{\prime},c^{\prime}\right\}  $
		& $\left\{  a^{\prime\prime},b^{\prime\prime}\right\}  $\\\hline
		$\left\{  c\right\}  $ & $\left\{  a^{\prime},b^{\prime}\right\}  $ &
		$\left\{  a^{\prime\prime},c^{\prime\prime}\right\}  $\\\hline
		$\emptyset$ & $\emptyset$ & $\emptyset$\\\hline
	\end{tabular}

	Table 6: ket table represent same abstract ket (a row) in three different bases
\end{center}

To see, for instance, that the $U^{\prime}$ vectors form a basis, they can be
added to yield the $U$-basis:

\begin{center}
	$\left\{  a^{\prime}\right\}  +\left\{  b^{\prime}\right\}  =\left\{
	a,b\right\}  +\left\{  a,b,c\right\}  =\left\{  c\right\}  $;
	
	$\left\{  a^{\prime}\right\}  +\left\{  b^{\prime}\right\}  +\left\{
	c^{\prime}\right\}  =\left\{  a,b\right\}  +\left\{  a,b,c\right\}  +\left\{
	b,c\right\}  =\left\{  b\right\}  $;
	
	$\left\{  b^{\prime}\right\}  +\left\{  c^{\prime}\right\}  =\left\{
	a,b,c\right\}  +\left\{  b,c\right\}  =\left\{  a\right\}  $.
\end{center}

\subsection{Analysis of the double-slit experiment}\label{subsec13}

A super-simple model of the two-slit experiment that captures the essential
ideas can be constructed using only $U=\left\{  a,b,c\right\}  $. In the setup
for the experiment, $U=\left\{  a,b,c\right\}  $ represent three vertical
distances. There is a particle emitter at $\left\{  b\right\}  $, the two-slit
screen with the slits at $\left\{  a\right\}  $ and $\left\{  c\right\}  $,
then the detection wall. The particle gun, the two-slit screen, and the
detection wall are equally spaced horizontally to represent the distance a
particle travels in one time unit.

\begin{figure}[h]
	\centering
	\includegraphics[width=0.7\linewidth]{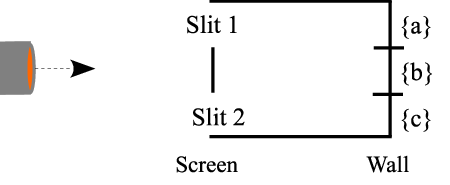}
	\caption{The setup for the two-slit experiment}
	\label{fig:two-slit-setup}
\end{figure}

The mathematical setting is $\wp\left(  U\right)  \cong \mathbb{Z}_{2}^{3}$ but there are no inner products in vector spaces over the finite
fields so the type II evolution is not one that takes an orthonormal basis
into an orthonormal basis (i.e., a unitary transformation) but simply a
non-singular linear transformation that takes a basis to a basis. In this
case, we choose the transformation of the $U$-basis to the $U^{\prime}$-basis.
This mimics the dynamics of a wave spreading out since: $\left\{  a\right\}
\longmapsto\left\{  a^{\prime}\right\}  =\left\{  a,b\right\}  $, $\left\{
b\right\}  \longmapsto\left\{  b^{\prime}\right\}  =\left\{  a,b,c\right\}  $,
and $\left\{  c\right\}  \longmapsto\left\{  c^{\prime}\right\}  =\left\{
b,c\right\}  $.

A particle is emitted at $\left\{  b\right\}  $ and in one time unit, it
transforms into $\left\{  b^{\prime}\right\}  =\left\{  a,b,c\right\}  $. The
first state reduction is at the two-slit screen. If the particle hits at
$\left\{  b\right\}  $, then nothing further happens. If the particle does not
hit at $\left\{  b\right\}  $, then the state reduction is to the
superposition $\left\{  a,c\right\}  $ at the screen. With equiprobable
outcomes, the probability of the particle hitting at $\left\{  b\right\}  $ in
the two-slit screen is $\frac{1}{3}$ and the probability of the superposition
$\left\{  a,c\right\}  $ result is $\frac{2}{3}$. Then we have the two cases
depending on whether or not there is detection at the slits.

\textbf{Case 1}: \textbf{Detection at the slits}. Then the superposition
reduces to either the classical state $\left\{  a\right\}  $, i.e., goes
through slit 1, or the classical state $\left\{  c\right\}  $, i.e., goes
through slit 2.

\begin{quotation}
	\noindent If you could, in principle, distinguish the alternative final states
	(even though you do not bother to do so), the total, final probability is
	obtained by calculating the probability for each state (not the amplitude) and
	then adding them together. If you cannot distinguish the final states even in
	principle, then the probability amplitudes must be summed before taking the
	absolute square to find the actual probability. \cite[3-16]{feynman:v3mill-ed}
\end{quotation}

\noindent In terms of that Feynman rule for the addition of probabilities
\cite[pp. 110-111]{jaeger:qobjects}, the interaction of the superposition
$\left\{  a,c\right\}  $ with the detectors distinguished between the two
alternatives (going through slit 1 or through slit 2) so thereafter the
probability of getting from the emitter to any position on the detection wall
is calculated by adding the probabilities, e.g., from emitter to $\left\{
a\right\}  $ at the wall, from the emitter to $\left\{  b\right\}  $ at the
wall , and from the emitter to $\left\{  c\right\}  $ at the wall. The
probability of going from $\left\{  a\right\}  $ at the screen to $\left\{
a\right\}  $ or $\left\{  b\right\}  $ at the wall is $\frac{1}{2}$ each.
Similarly the probability of going from $\left\{  c\right\}  $ at the screen
to $\left\{  b\right\}  $ or $\left\{  c\right\}  $ at the wall is $\frac
{1}{2}$ each. Since the path from the emitter to $\left\{  a\right\}  $ at the
wall has two distinguishable subprocesses (from emitter to $\left\{
a\right\}  $ at the screen and from $\left\{  a\right\}  $ at the screen to
$\left\{  a\right\}  $ at the wall), then, by Feynman's rule for
multiplication of probabilities, their probabilities must be multiplied
\cite[p. 111]{jaeger:qobjects}. The probability from the emitter to $\left\{
a\right\}  $ at the screen is $\frac{2}{3}\frac{1}{2}=\frac{1}{3}$ and from
$\left\{  a\right\}  $ at the screen to $\left\{  a\right\}  $ at the wall is
$\frac{1}{2}$ so the probability from the emitter to $\left\{  a\right\}  $ at
the wall is $\frac{1}{3}\frac{1}{2}=\frac{1}{6}$. In the same manner, the
probability from the emitter to $\left\{  b\right\}  $ at the wall via
$\left\{  a\right\}  $ at the screen is also $\frac{1}{6}$. But the particle
can also arrive at $\left\{  b\right\}  $ at the wall by going through
$\left\{  c\right\}  $ at the screen so that is another $\frac{1}{6}$ for
$\left\{  b\right\}  $ at the wall. And $\left\{  c\right\}  $ at the wall can
only be reached from $\left\{  c\right\}  $ at the screen so the probability
from emitter to $\left\{  c\right\}  $ at the wall is $\frac{1}{6}$. Hence we
have the probability distribution at the wall given in Figure 10 where all the
probabilities sum to $1$.

\begin{figure}[h]
	\centering
	\includegraphics[width=0.7\linewidth]{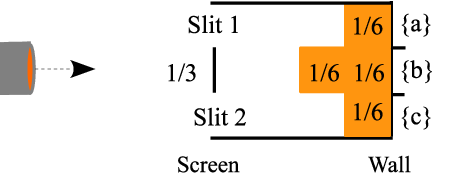}
	\caption{Bar chart of probabilities in Case 1 of detection at the slits}
	\label{fig:case1-bar-graph}
\end{figure}

These processes can all be traced in the partition lattice--remembering that
upward (dashed) arrows are state reductions due to distinctions being made,
and non-upward (solid) arrows are linear evolutions. The first step from the
emitter to the screen ($\left\{  b\right\}  \longmapsto\left\{  b^{\prime
}\right\}  =\left\{  a,b,c\right\}  $) and then the state reduction ($\left\{
a,b,c\right\}  \longmapsto\left\{  b\right\}  $ or $\left\{  a,c\right\}  $)
at the screen are depicted in Figure 11.

\begin{figure}[h]
	\centering
	\includegraphics[width=0.7\linewidth]{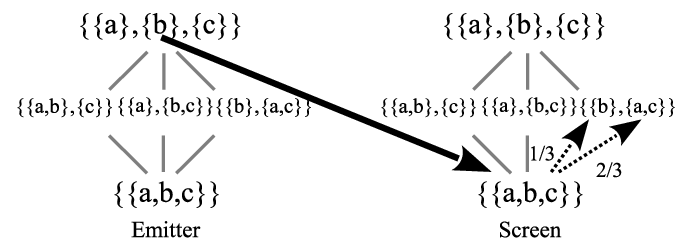}
	\caption{Linear evolution from emitter at $\left\{  b\right\}  $ to
		$\left\{  b^{\prime}\right\}  =\left\{  a,b,c\right\}  $ at screen and the
		state reduction by the screen}
	\label{fig:gun-to-screen-lattice}
\end{figure}

Then at the screen, the detectors distinguished the two parts of the
superposition $\left\{  a,c\right\}  $ and the two parts linearly evolve
separately to the wall (two solid arrows) which again causes the state
reductions indicated by the dashed upward arrows in Figure 12.

\begin{figure}[h]
	\centering
	\includegraphics[width=0.7\linewidth]{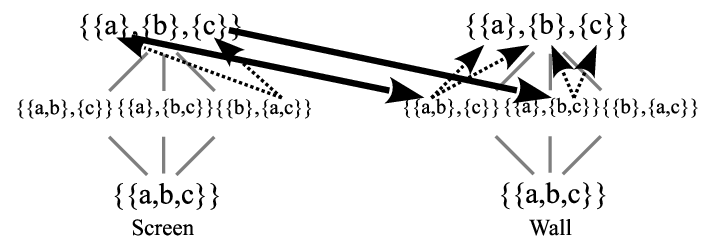}
	\caption{State reductions at the screen, linear evolutions to the wall, and
		state reductions at the wall}
	\label{fig:case1-screen-to-wall}
\end{figure}

\noindent The probabilities (all half-half in Figure 12) are attached only to
the upward dashed arrows and can be multiplied along the paths from emitter to
a distinguished hit at the wall.

\textbf{Case 2: No detection at the slits}. With no detection at the slits the
superposition $\left\{  a,c\right\}  $ at the screen continues to evolve
linearly at the quantum level since there was not distinguishing at the slits
to emerge into the classical states of going through slit 1 or going through
slit 2. The linear evolution of the superposition is:

\begin{center}
	$\left\{  a,c\right\}  =\left\{  a\right\}  +\left\{  c\right\}
	\longmapsto\left\{  a^{\prime}\right\}  +\left\{  c^{\prime}\right\}
	=\left\{  a,b\right\}  +\left\{  b,c\right\}  =\left\{  a,c\right\}  $
\end{center}

\noindent where the destructive interference occurred at the step $\left\{
a,b\right\}  +\left\{  b,c\right\}  =\left\{  a,c\right\}  $. The
superposition at the screen was not distinguished there so it never rose to
the classical level of going through one of the slits. The evolution and state
reduction to get from the emitter to the superposition is as before so
$\left\{  a,c\right\}  $ at the screen occurs with probability $\frac{2}{3}$.
Then the state evolves linearly to $\left\{  a,c\right\}  $ at the wall where
the states are distinguished to $\left\{  a\right\}  $ or $\left\{  c\right\}
$ with the probability $\frac{2}{3}\frac{1}{2}=\frac{1}{3}$ each. This give
the probability stripes characteristic of the interference effects and is
depicted in Figure 13. (where all the probabilities add to $1$).

\begin{figure}[h]
	\centering
	\includegraphics[width=0.7\linewidth]{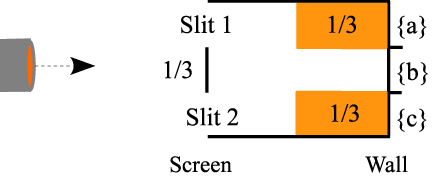}
	\caption{Case 2 of no detections so superposition evolves linearly with
		destructive interference at $\left\{  b\right\}  $}
	\label{fig:case2-bar-graph}
\end{figure}

The evolution up to the superposition $\left\{  a,c\right\}  $ at the screen
is as before so we can picture the remaining changes using the partition
lattices in Figure 14.

\begin{figure}[h]
	\centering
	\includegraphics[width=0.7\linewidth]{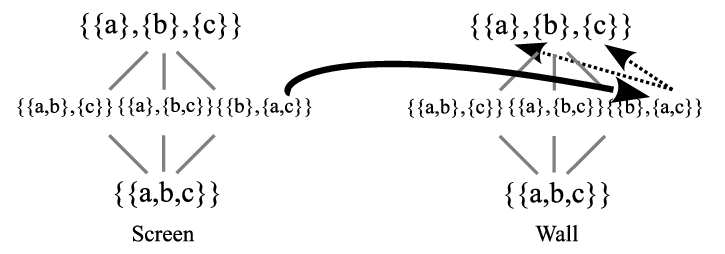}
	\caption{Linear evolution at the quantum level so classical events of going
		through slit 1 or 2 at screen did not occur}
	\label{fig:case2-screen-to-wall}
\end{figure}

In this manner, the key question: ``With no detection at the
slits, how does the particle get to the detection wall without (the classical
events of) going through slit 1 or going through slit 2, neither of which
would show interference?'' can be answered without resorting
the magical `explanation' of wave-particle complementarity.

\section{Closing the circle: state reduction = superposition$^{-1}$}\label{sec9}

\textit{State reduction is the inverse of superposition}. Superposition makes
definite states indefinite with various amplitudes, and state reduction is an
interaction that make the superposed alternatives distinct or distinguished
with various amplitudes (whose absolute squares give probabilities of the
distinct definite states). This relationship between superposition (making
indistinctions) and state reduction (making distinctions) closes the circle of
explanation by connecting the key non-classical notion of superposition and
the key mystery of state reduction (``collapse of the wave
packet'' etc.). We have no need to go deeper into just what a
superposition is or just what the distinguishing is in a state reduction since
quantum mechanics is a framework theory. Quantum field theory goes deeper into
quantum reality within that framework.

Hermann Weyl's use of a grating \cite[p. 255]{weyl:phil} to illustrate state
reductions provides an intuitive `pasta machine' treatment. Figure 15 has a
grating (on the left) of different pasta shapes. We think of the particle
going into the interaction with the pasta grating as being a ball of pasta
that is a superposition of the different shapes. The grating distinguishes
between the shapes in the superposition so the result is a definite shape.
Since the alternative paths through the grating are distinguished, the Feynman
rule specifies that the probability for each path is computed separately and
then added together to get the probability of going from A to B.

\begin{figure}[h]
	\centering
	\includegraphics[width=0.9\linewidth]{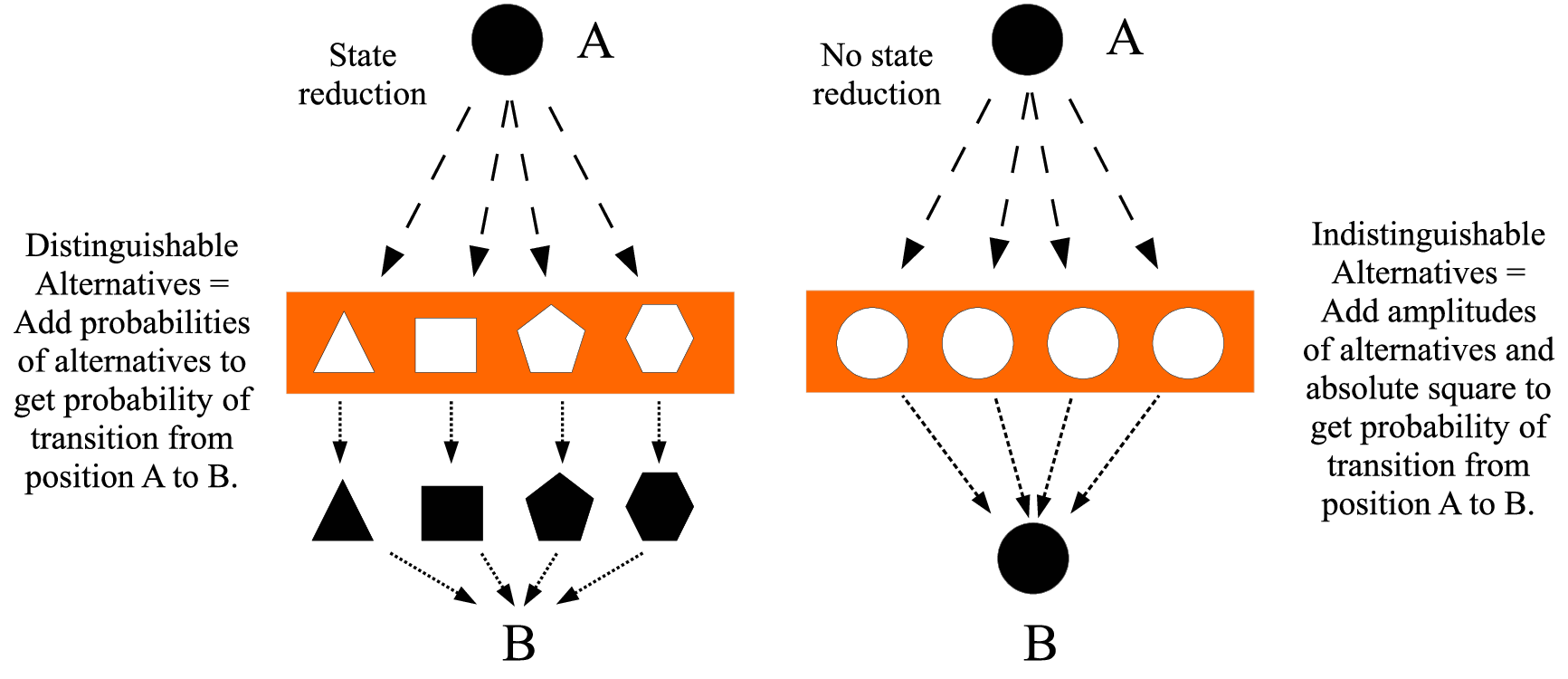}
	\caption{The two Feynman rules for a superposition having an interaction
		the does or does not distinguish the alternatives}
	\label{fig:two-gratings}
\end{figure}

On the right-hand side is the case of a `null grating' that does not
distinguish the alternative paths through the grating so there is no state
reduction caused by the grating and the path amplitudes are added and then the
absolute square taken to get the probability of going from A to B.

\section{Logical information in quantum mechanics}\label{sec10}

There is a long tradition that considers information and QM to be connected.
The connection is quite direct. Superposition makes indistinctions (with
various amplitudes), state reduction creates distinctions, and the notion of
logical entropy is the measure of classical and quantum distinctions.

The classical logical entropy of a probability distribution $p$ is $h\left(
p\right)  =1-\sum_{i=1}^{n}p_{i}^{2}$. The quantum logical entropy of a
density matrix $\rho$ is $h\left(  \rho\right)  =1-\operatorname*{tr}\left[
\rho^{2}\right]  $ \cite{ell:qic}. The interpretations are the probability
measures of distinctions:

\begin{itemize}
	\item $h\left(  p\right)  $ = the probability that two random draws from $U$
	give a distinction (of $\mathbf{1}_{U}$), and
	
	\item $h\left(  \rho\right)  $ = the probability that two independent
	measurements of the same prepared state $\rho$ give different eigenvalues.
\end{itemize}

Moreover, logical entropy measures measurement. The L\"{u}ders mixture
operation describes changes in density matrices with projection-valued
measurement (PVM): $\hat{\rho}=\sum_{r\in f\left(  U\right)  }P_{r}\rho P_{r}%
$. Each non-zero off-diagonal entry in $\rho$ represents a quantum
indistinction or coherence indicating that the corresponding diagonal states
are cohered together in the superposition.

\begin{quotation}
	\noindent For this reason, the off-diagonal terms of a density matrix ... are
	often called ``quantum coherences'' because
	they are responsible for the interference effects typical of quantum mechanics
	that are absent in classical dynamics. \cite[p. 177]{auletta:qm}
\end{quotation}

\noindent If in the projective measurement transition from $\rho$ to
$\hat{\rho}$, a quantum indistinction $\rho_{jk}\neq0$ is zeroed, i.e.,
$\rho_{jk}\rightsquigarrow0$, that means the corresponding eigenstates are
distinguished or decohered in the resulting mixed state. That increases the
distinctions so the quantum logical entropy goes up. That increase in
distinctions $h\left(  \hat{\rho}\right)  -h\left(  \rho\right)  $ is
precisely measured the sum of the absolute squares of those new distinctions
\cite{ell:qic},

\begin{center}
	$h\left(  \hat{\rho}\right)  -h\left(  \rho\right)  =\sum_{\rho_{jk}%
		\rightsquigarrow0}\left\vert \rho_{jk}\right\vert ^{2}$.
\end{center}

\section{Rota's twelve-fold way and quantum statistics}\label{sec11}

\subsection{Functions defined by elements and distinctions}\label{subsec14}

The basic dual notions of elements and distinctions can be used to give the
`natural' definition of a set function $f:X\rightarrow Y$ and the two special
cases of injective and surjective functions. Let $R\subseteq X\times Y$ be a
binary relation on $X\times Y$. Then consider the following definitions about
preserving (transmitting) or reflecting elements and distinctions.

\begin{itemize}
	\item $R$ is said to \textit{preserve elements} if for all $x\in X$, there is
	a $y\in Y$ such that $\left(  x,y\right)  \in R$.
	
	\item $R$ is said to \textit{reflect elements} if for all $y\in Y$, there is a
	$x\in X$ such that $\left(  x,y\right)  \in R$.
	
	\item $R$ is said to \textit{preserve distinctions} if for any $\left(
	x,y\right)  ,\left(  x^{\prime},y^{\prime}\right)  \in R$, if $x\neq
	x^{\prime}$, then $y\neq y^{\prime}$.
	
	\item $R$ is said to \textit{reflect distinctions} if for any $\left(
	x,y\right)  ,\left(  x^{\prime},y^{\prime}\right)  \in R$, if $y\neq
	y^{\prime}$, then $x\neq x^{\prime}$.
\end{itemize}

Ordinarily, a relation $R\subseteq X\times Y$ is said to define a function
$f:X\rightarrow Y$ if it is defined for all $x\in X$ and is single-valued. But
being defined on all $x\in X$ is the same as preserving (or transmitting)
elements and being single-valued is the same as reflecting distinctions. Hence
the natural elements-and-distinctions definition of a set \textit{function}
$f:X\rightarrow Y$ is a binary relation that preserves elements and reflects
distinctions. Moreover, the language of preserving indicates directionality in
the same direction and the language of reflecting indicates directionality in
the opposite direction so both parts of the function definition have the same
directionality. The other two special types of functions are the ones that
satisfy the other two conditions, i.e., injections preserve distinctions and
surjections reflect elements.

\subsection{Rota's twelve-fold way}\label{subsec15}

The language needed to understand quantum mechanics is the language of
indefiniteness versus definiteness or indistinguishability versus
distinguishability. That is also the language used to describe the different
enumerations of the ways of distributing $k$ balls (particles) into $n$ boxes
(states). Firstly, the balls can be distinguishable or indistinguishable,
where ``indistinguishable balls'' means that
a given distribution of balls to boxes does not change if the balls are
permuted. Secondly, the boxes can be distinguishable or indistinguishable,
where ``\ indistinguishable boxes'' means that
a given distribution does not change if the boxes are permuted. That gives $4$
different types of distribution of balls into boxes. But the functions taking
balls into boxes could have $3$ different types, arbitrary functions,
functions preserving distinctions (injections), or functions reflecting
elements (surjections). Hence gives $3\times4=12$ different ways to distribute
balls into boxes which gives Rota's ``twelve-fold
way'' \cite[p. 71]{stanley;enum-comb-1}.

\subsection{Maxwell-Boltzmann distribution}\label{subsec16}

The conceptual language of enumerative combinatorics connects to the
description of the worldview of classical physics as ``
definite all the way down.'' Hence in the classical case, the
balls or particles are distinguishable, the boxes or states are
distinguishable, and the distributions is by arbitrary functions. That yields
the Maxwell-Boltzmann (MB) distribution.

There are $k!$ permutations of the distinguishable particles into the $n$
states with the occupation numbers $\theta_{1},...,\theta_{n}$ which sum to
$k$. How many distributions are there with those occupation numbers? It is
perhaps surprising that the occupation numbers don't matter so long as they
sum to $k$ since the vertical separator lines dividing the $k!$ permutations
can be moved arbitrarily in the following Figure 16.

\begin{figure}[h]
	\centering
	\includegraphics[width=0.7\linewidth]{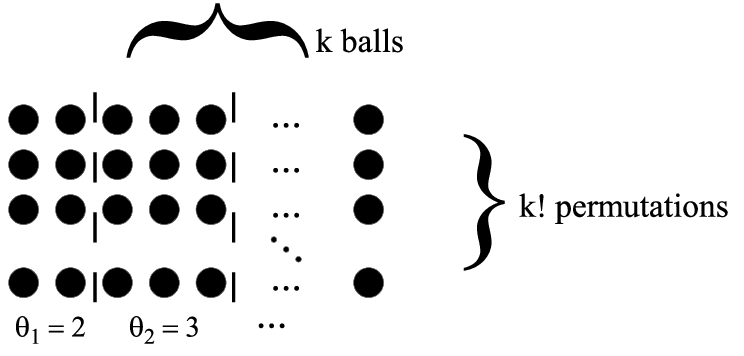}
	\caption{$k!$ permutations = \# of distributions between boxes with any
		$\theta_{1},...,\theta_{n}$ (summing to $k$)}
	\label{fig:k-factorial-permutations}
\end{figure}

\noindent What counts is which $\theta_{i}$ distinguishable balls are in the
$i^{th}$ distinguishable box but there is such thing as an ordering of the
balls in the box so the $k!$ permutations need to be divided by the
$\theta_{i}!$ for $i=1,...,n$ to obtain the total number of distributions
which is the well-known \textit{multinomial coefficient}:

\begin{center}
	$\binom{k}{\theta_{1},...,\theta_{n}}=\frac{k!}{\theta_{1}!...\theta_{n}!}$.
\end{center}

\noindent Since the distribution of balls to boxes is by arbitrary functions,
there are $n^{k}$ functions from balls to boxes and each one is considered
equally probable so the probability of distributing the $k$ balls into the
boxes with the occupation numbers $\theta_{1},...,\theta_{n}$ in the
\textit{Maxwell-Boltzmann statistics} \cite[p. 39]{feller:vol1} is:

\begin{center}
	$\frac{k!}{\theta_{1}!...\theta_{n}!}/n^{k}$.
\end{center}

\subsection{Fermi-Dirac distribution}\label{subsec17}

The MB distribution is appropriate for classical physics which is definite all
the way down. But quantum physics is only definite down to a certain level,
e.g., determined by a CSCO. The premise in Leibniz's Principle of Identity of
Indistinguishables is that reality is definite all the way down. Given two
allegedly different classical particles, if they are indistinguishable all the
way down, then they are the same particle. If reality is definite only down to
a certain level, then one should two types of particles. Some particles
(fermions) might be like classical particles in that if two allegedly
different particles were indistinguishable down to the given level, then they
are unique, i.e., are the same particle. But now there might be another type
of particle (bosons) where indistinguishability down to the given level is
insufficient to uniquely determine the particle so there might be many
numerically distinct such particles that are indistinguishable and in the same state.

As an analogy, consider a package delivery system where the address was only
definite down to the level of street number, i.e., country, city, city code,
and street name and number but no further definiteness. Then two cases arise,
namely where that was sufficient to uniquely determine the addressee and it
was insufficient to determine the addressee. It would be sufficient in the
case of a neighborhood zoned for single-family residences, i.e., a
``fermionic'' neighborhood. In how many ways
could Santa Claus deliver at most one of $k$ standard (indistinguishable)
Christmas packages among $n$ (distinguishable) street-number addresses?

In terms of the packages to street-number address mapping, it is
distinction-preserving since different packages must go to different street
numbers. Hence in terms of Rota's twelve-fold way, it is the number of
injective functions mapping indistinguishable balls into distinguishable
boxes. That is given by the number of ways to distribute $k$ balls into $n$
boxes with at most one ball per box which is the binomial coefficient:

\begin{center}
	$\binom{n}{k}=\frac{n!}{k!\left(  n-k\right)  !}=\frac{n\left(  n-1\right)
		...\left(  n-k+1\right)  }{k!}=\frac{n^{\underline{k}}}{k!}$
\end{center}

\noindent where $n^{\underline{k}}=n\left(  n-1\right)  ...\left(
n-k+1\right)  $ is the \textit{falling factorial}. If each distribution of at
most one ball per box is equiprobable, the F\textit{ermi-Dirac (FD)}
\textit{distribution} gives the probability of each distribution as
$1/\binom{n}{k}$.

\subsection{Bose-Einstein distribution}\label{subsec18}

In our package address metaphor, the alternative is the sort of neighborhood
where an address definite only down to the street number is insufficient to
determine an addressee, e.g., a neighborhood zoned to allow multifamily houses
or apartment buildings, i.e., a ``bosonic''%
\ neighborhood. In the enumeration for the Fermi-Dirac distribution, each ball
was seen as occupying a box so no more balls could be put in the box. The
alternative of allowing many balls in a box (e.g., many addressees at a
street-number address), is to see each ball in a box as allowing another ball
on either side of it as illustrated in Figure 17.

\begin{figure}[h]
	\centering
	\includegraphics[width=0.7\linewidth]{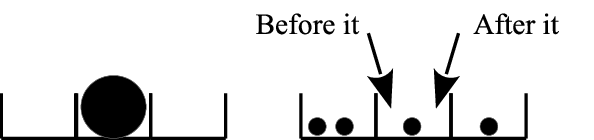}
	\caption{Illustration of FD restriction of at most one ball in a box versus
		allowing repetition of balls in a box}
	\label{fig:fd-be-pix}
\end{figure}

\noindent In the case of allowing repeated balls in a box, the addition of a
ball could be seen as creating two more landing places for balls, namely
before and after it in the box. Hence the total number of landing places is
computed as the \textit{rising factorial}:

\begin{center}
	$n^{\overline{k}}=n\left(  n+1\right)  ...(n+k-1)$.
\end{center}

\noindent The balls are indistinguishable so the total number of combinations
with repetitions is:

\begin{center}
	$\frac{n^{\overline{k}}}{k!}=\frac{n\left(  n+1\right)  ...(n+k-1)}{k!}%
	=\frac{\left(  n+k-1\right)  !}{k!\left(  n-1\right)  !}=\binom{n+k-1}%
	{k}=\binom{n+k-1}{n-1}$.
\end{center}

\noindent Each such combination is considered equiprobable in the
\textit{Bose-Einstein distribution} so the probability distribution is given
by the reciprocal.

In terms of Rota's twelve-fold way, the Bose-Einstein case is
indistinguishable balls, distinguishable boxes, and the distribution of balls
to boxes by arbitrary functions, instead of by distinction-preserving or
injective functions as in the Fermi-Dirac case. The Fermi-Dirac distribution
gives the number of combinations without repetition of balls in a box compares
to the Bose-Einstein distribution of allowing repetitions by using the falling
or rising factorial in the formulas:

\begin{center}
	FD: $\frac{n^{\underline{k}}}{k!}$ or BE: $\frac{n^{\overline{k}}}{k!}$.
	
	The Rota way to quantum statistics
\end{center}

The duality in combinatorial theory \cite[p. 295]{stanley:enum-comb-2} of the
two enumerations as not allowing repetitions for ``
fermionic'' balls and allowing repetitions for
``bosonic'' balls highlights the Pauli
exclusion principle for fermions.

\section{Concluding remarks}\label{sec12}

This journey to understanding quantum mechanics started at the logical level
of developing the logic of partitions dual to the usual Boolean logic of
subsets. The notion of distinctions of partition rose to the foreground as
having the analogous dual role to elements of subsets. Since logical
probability starts with the (normalized) quantitative element-count:
$\Pr\left(  S\right)  =\frac{\left\vert S\right\vert }{\left\vert U\right\vert
}$. so logical information starts with the (normalized) quantitative
distinction-count: $h\left(  \pi\right)  =\frac{\left\vert \operatorname*{dit}%
	\left(  \pi\right)  \right\vert }{\left\vert U\times U\right\vert }$ which
measures information-as-distinctions by logical entropy. This part of the
journey was guided by the intuitions of Gian-Carlo Rota.

At that point, it was surmised that there is a fundamental `metaphysical'
duality based on dealing with single elements or pairs of elements.

\begin{itemize}
	\item For a single element, the basic question is `to be or not to be' in a
	subset or `to have or not have' a property.
	
	\item For a pair of elements, the basic question is `same or different,'
	`indistinct or distinct,' `inequivalent or equivalent,' `distinguishable or
	indistinguishable,' or `identity or difference.'
\end{itemize}

\noindent This duality is brought out clearly and extensively developed in
category theory. The duality extends to physical theories. Classical physics
is the physics of particles that are `definite all the way down' so a particle
will ultimately have a property or its negation apply to it. In contrast,
quantum physics is characterized by not being `definite all the way down',
i.e., not being any more definite than a characterization given by a complete
set of commuting observables. The logic of indefiniteness (within blocks) and
definiteness (between blocks) is the logic of partitions. A simplified model
that preserves the essentials is given by a lattice of partitions that goes
from a pure state at the bottom to the corresponding completely decomposed
mixed state at the top with all the mixed states obtainable by projective
measurement in between.

A key step in completing the circle of explanation was seeing the inverse
connection between the quantum interpretation of superposition (making the
definite states indistinct with various amplitudes) and the Feynman treatment
of state reduction (distinguishing the superposed alternatives with various amplitudes).

Gian-Carlo Rota was the father of modern combinatorial theory. Rota's
twelve-fold way is based on the concepts of distinguishability or
indistinguishability from the partition side of the fundamental duality, and
it was used to highlight the difference between fermions and bosons in QM.
Quantum physics is only definite down to a certain level as determined by a
complete set of commuting observables (CSCO) so, intuitively, one would expect
two types of particles; those where a CSCO-description is enough to uniquely
designate a particle (fermions) and those where such a description is
insufficient (bosons). Classical physics is definite all the way down, so
there is only one type of classical particle which is uniquely designated by
`going down far enough.' That is, by Leibniz's PII, given two allegedly
different classical particles, either by going down far enough there is an
attribute that applies to one and not the other--or otherwise they are
identical, i.e., there is only one particle with that total description. In
that sense, Leibniz's PII should be seen as the ``Pauli
exclusion principle'' for a physical reality that is definite
all the way down. When reality is not definite all the way down, then some
particles may still satisfy that principle (fermions) but others may not
(bosons). Weyl clearly brought out that connection.

\begin{quotation}
	\noindent The upshot of it all is that the electrons satisfy Leibniz's
	\textit{principium identitatis indiscernibilium}, or that the electronic gas
	is a ``monomial aggregate'' (Fermi-Dirac statistics). ... As to the
	Leibniz-Pauli exclusion principle, it is found to hold for electrons but not
	for photons. \cite[p. 247]{weyl:phil}
\end{quotation}

The lattice of partitions illustrates (at the skeletal level of support sets)
that reality is composed of the classical part of definiteness and the quantum
underworld of indefiniteness, the iceberg metaphor being appropriate for the
two parts of reality which Heisenberg, Shimony, and many others described as
the actual and potential (indefinite) parts of reality. Shimony described this
vision of QM as the ``Literal Interpretation'' based on the notions of objective indefiniteness, objective chance, and the
quantum state being a complete description.


\backmatter



\section*{Declarations}

\begin{itemize}
\item No funding to report.
\item There are no conflict of interest or competing interests.
\item Ethics approval and consent to participate. Not applicable.
\item Consent for publication. Not applicable.
\item Data availability. Not applicable.
\item Materials availability. Not applicable.
\item Code availability, Not applicable.
\item Author contribution. Not applicable.
\end{itemize}

\end{document}